\documentclass{jfm}
\usepackage{graphicx}
\usepackage{epstopdf, epsfig}
\usepackage{subfigure}
\usepackage{amsmath}
\usepackage{amssymb}
\usepackage{amsfonts}
\usepackage{latexsym}
\usepackage{empheq}
\usepackage{cases}
\usepackage{xcolor}
\usepackage{mathtools}

\shorttitle{Axisymmetric screech-tone modelling}
\shortauthor{M. Mancinelli, V. Jaunet, P. Jordan, A. Towne}

\title{A complex-valued resonance model for axisymmetric screech tones in supersonic jets}

\author{Matteo Mancinelli\aff{1}
  \corresp{\email{matteo.mancinelli@univ-poitiers.fr}},
  Vincent Jaunet\aff{1},
  Peter Jordan\aff{1},
 \and Aaron Towne\aff{2}}

\affiliation{\aff{1}D\'{e}partement Fluides Thermique et Combustion, Institut Pprime - CNRS-Universit\'{e} de Poitiers-ENSMA, 11 Boulevard Marie et Pierre Curie, 86962 Chasseneuil-du-Poitou, Poitiers, France
\aff{2}Department of Mechanical Engineering, University of Michigan, 2350 Hayward Street, Ann Arbor, MI 48109, USA}

\begin{document}

\maketitle

\begin{abstract}
We model the resonance mechanism underpinning generation of A1 and A2 screech tones in an under-expanded supersonic jet. Starting from the resonance model recently proposed by \cite{mancinelli2019screech}, where the upstream-travelling wave is a neutrally-stable guided jet mode, we here present a more complete linear-stability-based model for screech prediction. We study temperature and shear-layer thickness effects and show that, in order to accurately describe the experimental data, the effect of the finite thickness of the shear layer must be incorporated in the jet-dynamics model. We then present an improved resonance model for screech-frequency predictions in which both downstream- and upstream-travelling waves may have complex wavenumber and frequency. This resonance model requires knowledge of the reflection coefficients at the upstream and downstream locations of the resonance loop. We explore the effect of the reflection coefficients on the resonance model and propose an approach for their identification. The complex-mode model identifies limited regions of frequency-flow parameter space for which the resonance loop is amplified in time, a necessary condition for the resonance to be sustained. This model provides an improved description of the experimental measurements. 
\end{abstract}

\begin{keywords}
Authors should not enter keywords on the manuscript, as these must be chosen by the author during the online submission process and will then be added during the typesetting process (see http://journals.cambridge.org/data/\linebreak[3]relatedlink/jfm-\linebreak[3]keywords.pdf for the full list)
\end{keywords}

\section{Introduction}
\label{sec:intro}
The modelling of screech in imperfectly expanded supersonic jets is a long-standing problem to which a large body of work has been devoted (for comprehensive reviews on the subject see \cite{raman1999supersonic} and \cite{edgington2019aeroacoustic}). It is an accepted fact that screech arises when coherent turbulent structures traverse the shock-cells of imperfectly expanded supersonic jets, following which two things may occur: (1) a directive sound field may set up with peak radiation oriented in the upstream direction; (2) resonance may be established. If resonance is established, the directive sound field is tonal. If resonance is not established, the directive sound field exists but is broadband. There are therefore two mechanisms that are interesting to consider: the sound generation mechanism; and the closure mechanism that enables resonance.

\subsection{Sound generation mechanism}
\label{subsec:intro_sound}
\cite{powell1953mechanism} considered a simplified model of the interaction between coherent structures and shock-cells, representing this as a distribution of point (monopole) sources that fluctuate with a phase relationship that leads to directive radiation in the upstream direction. This phased-array model has been widely used since then, for instance in the work of \cite{harper1974noise}. An alternative model was proposed by \cite{tam1982shock} and \cite{tam1986proposed}, where the rational is developed in wavenumber space. Downstream-travelling Kelvin-Helmholtz waves, of wavenumber $k_{KH}^+$, interact non-linearly with a stationary shock-cell structure characterised by wavenumber $k_{s}$, and the interaction leads to a difference wavenumber $k_p^-=k_{KH}^+-k_{s}$. Sound is then generated by the so-called Mach-wave mechanism, which radiates in the upstream direction because of the negative phase-speed (more recent works by \cite{towne2017acoustic} and \cite{edgington2021waves} have established that the group velocity of this wave is also negative). As shown by \cite{lele2005phased}, the phased-array and wave-interaction models for sound radiation are equivalent, and there is a solid body of work suggesting that the sound-generation mechanisms of both BBSAN and screech are indeed underpinned by this distributed-source behaviour (see among many \cite{tam1982shock, tam1986proposed, shen2002three, wong2021wavepacket}).

An alternative sound-generation mechanism, known as shock leakage, has been proposed by \cite{manning1998numerical} and \cite{suzuki_lele_2003}. Shock leakage occurs when the passage of coherent structures across a shock leads to a time-varying interruption of the total-internal reflection by which shocks are contained in the flow. The waves that were trapped in the jet are released and propagate with an upstream-oriented directivity. Whilst sound emissions from multiple sources have been experimentally observed (see e.g. \cite{edgington2021generation}), this mechanism can be in principle directive without requiring a streamwise distribution of sources. It remains to be established whether the directive sound beam associated with BBSAN and screech is due to the phased-array (Mach wave) source effect, or the shock-leakage mechanism, or both.

\subsection{Resonance closure mechanism}
\label{subsec:intro_closure}
As outlined above, the sound radiation may be broadband or tonal. Which it is depends on whether resonance is established. Resonance occurs when a phase relationship is established between downstream- and upstream-travelling disturbances such that these waves reinforce one another. A simple model for this, in terms of frequency selection, was provided by \cite{powell1953mechanism},

\begin{equation}
\frac{p-\phi^*}{f}=\frac{L_s}{\vert U^+\vert}+\frac{L_s}{\vert U^-\vert}\mathrm{,}
\label{eq:powell}
\end{equation}

\noindent where $U^+$ and $U^-$ are, respectively, the phase speeds of the downstream- and upstream-travelling disturbances, $f$ is the screech frequency, $p=N^++N^-$ is an integer corresponding to the number of wave cycles in the resonance loop, that is the sum of downstream- and upstream-travelling wave cycles, and $\phi^*$ is a constant related to the phase between the waves. $L_s$ is the distance between the points at which the two waves exchange energy, and for a simple line-distribution of point sources, such as considered by \cite{powell1953mechanism} or \cite{harper1974noise}, it is best expressed as

\begin{equation}
L_s=N_s\lambda_{s}=\frac{2\pi N_s}{k_s}\mathrm{,}
\label{eq:equispace}
\end{equation}

\noindent where $\lambda_{s}$ is the shock-cell spacing and $N_s$ is the number of the shock cells within the distance $L_s$. Equation \eqref{eq:powell} shows that, for a model in which the equivalent sources are assumed to be equi-spaced, frequency selection is independent of the number of equivalent point sources considered. This can be shown by substituting \eqref{eq:equispace} into \eqref{eq:powell}, thus implying the following prediction formula for the frequency,

\begin{equation}
f=\frac{p-\phi^*}{N_s}\frac{k_s}{2\pi}\left(\frac{U^+U^-}{U^--U^+}\right)\mathrm{.}
\end{equation}

Whether one considers a single shock-cell, or a series of shock cells, the model leads to the same frequency selection since the ratio of the difference between the number of wave cycles $p$ and the phase $\phi^*$ to the number of shock cells $N_s$ would remain unchanged. An equivalent formulation of the frequency selection criterion \eqref{eq:powell}, written in terms of wavenumber difference, is

\begin{equation}
\left(k_r^+ -k_r^-\right)L_s +\phi=2p\pi\mathrm{,}
\label{eq:powell_wave}
\end{equation}

\noindent where $k_r^\pm=2\pi f/U^\pm$ are the real parts of the wavenumbers associated with the downstream-travelling $k_{KH}^+$ and upstream-travelling $k_p^-$ waves and $\phi=2\pi\phi^*$ is the phase between the waves. It is possible to show that for an equi-spaced distribution of sources,

\begin{equation}
\frac{p-\phi^*}{N_s}=\text{constant} \qquad \xRightarrow{\text{implies}} \qquad \frac{k_r^+-k_r^-}{k_s}=\text{constant}\mathrm{,}
\end{equation}

\noindent thus linking the formulation presented by \cite{powell1953mechanism} and expressed in \eqref{eq:powell} and \eqref{eq:powell_wave} to the wave-interaction model first proposed by \cite{tam1982shock}.

But shock-containing jets differ from this simplified picture in two important ways: (1) the shock-cell spacing is not constant, and (2) the downstream-travelling wave is convectively unstable. The first of these observations alone points to a competition between equivalent sources in selecting the resonance frequency: the equations show how the individual equivalent sources $s$ will each try to select a different frequency, associated with the different distance $L_s$ of that source from the nozzle. The fact that screech occurs at a single frequency suggests that one of the equivalent sources wins the frequency selection competition. The second observation (spatially growing $k^+$ wave) shows how this competition may be won or lost: because of the growing amplitude of the downstream-travelling wave, reflection amplitudes (or, alternatively, equivalent source strenghts) will vary with streamwise position; it is likely that the strongest of these interactions is the one that will win the competition. In order to take this second aspect of the problem into account, a more complete resonance criterion model must be considered in which the growth rates of the waves are considered. Such a model has recently been used to describe jet-edge interaction by \cite{jordan2018jet}. The model is

\begin{subequations}
\begin{align}
& e^{\Delta k_i L_s}=\vert R_1R_2\vert\mathrm{,}\\
& \Delta k_r L_s +\phi =2p\pi\mathrm{,}
\end{align}
\end{subequations}

\noindent where $k_i$ refers to the imaginary part of the wavenumber, and $R_1$ and $R_2$ are reflection coefficients at the points where the waves exchange energy. Once again, in a model where equivalent sources are equispaced, frequency selection will be independent of the number of source points considered. Resonance strength, on the other hand, will depend on the number of equivalent sources, as each additional source will allow more energy to be fed back to the nozzle. However, when equivalent sources are not equi-spaced, competition occurs, and the equivalent source most likely to be chosen is that which produces the largest reflected amplitude (or equivalent source strength).

The screech-frequency prediction model we propose is based on this idea. Because of the non-equi-spaced character of equivalent sources in a spatially spreading, imperfectly expanded jet, and the spatial growth of the downstream-travelling wave, resonance closure is considered to be underpinned by a dominant reflection point (a dominant equivalent source). It is this equivalent source that determines frequency and growth rate of resonance.

\subsection{Upstream-travelling waves in the resonance loop and implications for frequency predictions}
\label{subsec:intro_predictions}
As shown by \cite{powell1992observations}, the staging evolution of the screech frequency with jet Mach number is characterised by switching between modes or stages: axisymmetric A1 and A2 modes \citep{merle1957nouvelles}, flapping B and D modes \citep{mercier2017experimental} and helical C mode \citep{edgington2014coherent}. Since the work of \cite{powell1953mechanism}, it has been assumed that the closure mechanism is provided by upstream-travelling free-stream acoustic waves. On the basis of this phenomenological description, several screech-frequency prediction models have been proposed \citep{powell1953mechanism, tam1986proposed, panda1999experimental, gao2010multi}. These provide only rough agreement with experimental data, and many screech features, such as staging, are not satisfactorily captured.

The idea that screech may not be closed by free-stream acoustic waves was first suggested by \cite{shen2002three}, who claimed that A1 and B screech modes involve free-stream acoustic waves, whereas A2 and C modes involve upstream-travelling guided jet modes. These guided modes were first studied by \cite{tam1989three} and have since been used to explain many resonance phenomena: in subsonic and supersonic impinging jets \citep{tam1990theoretical, bogey2017feedback}, in high-speed subsonic jets \citep{towne2017acoustic} and in a jet-flap interaction configuration \citep{jordan2018jet}. Most importantly, they have been shown to be active in screeching supersonic jets \citep{edgington2018upstream, gojon2018oscillation, edgington2021waves}. Given this insight, we recently developed a screech-frequency prediction model based on a resonance between downstream-travelling Kelvin-Helmholtz (K-H) waves and upstream-travelling guided jet modes \citep{mancinelli2019screech}. The phase speeds of both the K-H and upstream-travelling jet waves were provided by a cylindrical vortex-sheet (V-S) model. Contrary to the assertion of \cite{shen2002three}, the study showed that both A1 and A2 screech modes are underpinned by a resonance involving the upstream-travelling guided jet modes and screech-frequency predictions provided better agreement with experiments than does the classical prediction approach proposed by \cite{powell1953mechanism} using free-stream acoustic waves.

The model of \cite{mancinelli2019screech} considers an isothermal jet. We here consider the effect of jet temperature and show that a finite-thickness model is required to capture the temperature effects on screech generation. In the model of \cite{mancinelli2019screech}, consistent with what is frequently assumed in fluid-mechanics resonance phenomena (see e.g. \cite{rossiter1964wind} and \cite{rowley2002self} for resonance in cavity flows), both the upstream- and downstream-travelling waves are considered to be neutrally stable. While predictions provide good agreement with data, the simplification misses certain aspects of the observed behaviour: the model predicts tones that are not observed experimentally, for instance. This inaccuracy arises for both the free-stream- and guided-jet-mode-based resonance models.

We here explore the limitations of the neutral-mode approach presented in \cite{mancinelli2019screech} and consider a more complete model in which both frequency and wavenumber may be complex so that the growth rate of the waves involved in the resonance is included. This requires consideration of the upstream and downstream reflection mechanisms, which take form of a reflection-coefficient product. While the reflection coefficients could be determined by using a Wiener-Hopf method \citep{noble1958methods, rienstra2007acoustic} or by detailed numerical analysis, we here follow the approach used by \cite{jordan2018jet} and \cite{mancinelli2019reflection} to study resonance in a jet-flap interaction configuration and screeching jets; that is, we treat the reflection-coefficient product as a parameter and explore the predictions obtained by imposing its amplitude and phase. We then consider a simplified real-frequency-based variant of the model to educe values of the reflection-coefficient product from experimental data. On the basis of these results, we propose a functional form for the frequency-Mach-number dependence of the reflection-coefficient product. The refined resonance model, which includes finite-thickness and temperature effects contrary to that presented in \cite{mancinelli2019reflection}, provides a more complete description of the experimental observations, and, in particular, removes spurious tone predictions that occur with real-frequency models.

The paper is organised as follows. The resonance model is presented in \S\ref{sec:model}. \S\ref{sec:setup} describes the experimental set-up and instrumentation. Main results concerning the screech-frequency predictions and the estimation of the reflection-coefficient product are reported in \S\ref{sec:results}. Conclusions are finally presented in \S\ref{sec:conclusions}.

\section{Resonance models}
\label{sec:model}
In this section, we present the model based on a resonance between downstream-travelling K-H instability waves and two kinds of upstream-travelling wave: (i) free-stream acoustic waves, which are used in the usual screech scenario, and (ii) guided jet modes. Following \cite{towne2017acoustic} and \cite{mancinelli2019screech}, we use the terms downstream- and upstream-travelling to designate the direction of the energy transfer \citep{briggs1964electron, bers1983space}. Accordingly, the downstream- and upstream-travelling waves are denoted using superscripts $+$ and $-$, respectively. The Kelvin-Helmholtz instability, the guided jet modes and the free-stream acoustic waves are denoted $k_{KH}$, $k_p$ and $k_a$, respectively. Modes are obtained using linear stability theory.

\subsection{Jet models}
\label{subsec:jet_modelling}
The jet is modelled using locally parallel linear stability theory. All variables are normalised by the nozzle diameter $D$, and the ambient density and speed of sound $\rho_\infty$ and $c_\infty$, respectively. The Reynolds decomposition

\begin{equation}
q\left(x,r,\theta,t\right)=\overline{q}\left(r\right)+q'\left(x,r,\theta,t\right)
\label{eq:Re_decomposition}
\end{equation}

\noindent is applied to the flow-state vector $q$, where the mean and fluctuating components are $\overline{q}$ and $q'$, respectively. We assume the normal-mode ansatz,

\begin{equation}
q'\left(x,r,\theta,t\right)=\hat{q}\left(r\right)e^{i\left(kx+m\theta-\omega t\right)}\mathrm{,}
\label{eq:normal_ansatz}
\end{equation}

\noindent where $k$ is the streamwise wavenumber normalised by the nozzle diameter $D$, $m$ is the azimuthal mode and $\omega=2\pi StM_a$ is the non-dimensional frequency, where $St=fD/U_j$ is the nozzle-diameter-based Strouhal number and $M_a=U_j/c_\infty$ is the acoustic Mach number. As discussed by \cite{tam1990theoretical} and \cite{towne2017acoustic}, $k_p$ modes belong to a hierarchical family of waves identified by their azimuthal and radial orders $m$ and $n$, respectively, and exist as propagative waves only in a well defined $St$ number range (see \S\ref{subsec:real_analysis}). We here restrict attention to azimuthal mode $m=0$ due to the axisymmetry property of screech modes A1 and A2 and we let the radial order vary in the range $n=1,2$, which is sufficient for the frequency range of the screech modes of interest.

This theory is applied to two different models: a simplified cylindrical vortex sheet and a finite-thickness flow model. Both are governed by the compressible linearised Euler equations (LEE).

\subsubsection{Cylindrical vortex-sheet model}
\label{subsubsec:VS}
Following our previous work \citep{mancinelli2019screech}, the linear wave dynamics are first modelled using a cylindrical vortex sheet \citep{lessen1965inviscid, michalke1970note}. The vortex-sheet dispersion relation is

\begin{equation}
D\left(k,\omega;M_a,T,m\right)=\frac{1}{\left(1-\frac{kM_a}{\omega}\right)^2} + \frac{1}{T}\frac{I_m\left(\frac{\gamma_i}{2}\right)\left(\frac{\gamma_o}{2}K_{m-1}\left(\frac{\gamma_o}{2}\right) + mK_m\left(\frac{\gamma_o}{2}\right)\right)}{K_m\left(\frac{\gamma_o}{2}\right)\left(\frac{\gamma_i}{2}I_{m-1}\left(\frac{\gamma_i}{2}\right) - mI_m\left(\frac{\gamma_i}{2}\right)\right)}=0
\label{eq:dispersion}
\end{equation}

\noindent with

\begin{subequations}
\begin{align}
&\gamma_i = \sqrt{k^2-\frac{1}{T}\left(\omega-M_ak\right)^2}\mathrm{,}\label{eq:gamma_i}\\
&\gamma_o = \sqrt{k^2-\omega^2}\mathrm{,}\label{eq:gamma_o}
\end{align}
\end{subequations}

\noindent where $I_m$ and $K_m$ are modified Bessel functions of the first and second kind, respectively, and $T=T_j/T_\infty$ is the jet-to-ambient temperature ratio. The relation between the acoustic and jet Mach numbers is given by $M_j=U_j/c_j=M_a/\sqrt{T}$. The branch cut of the square root in \eqref{eq:gamma_i} and \eqref{eq:gamma_o} is chosen such that $-\pi/2\leq \rm{arg}\left(\gamma_{i,o}\right)<\pi/2$.

Frequency/wavenumber pairs $\left(\omega, k\right)$ that satisfy \eqref{eq:dispersion} define eigenmodes of the vortex sheet for given values of $m$, $M_a$ and $T$. To find these pairs, we specify a real or complex frequency $\omega$ and compute the associated eigenvalues $k$ for the K-H and guided jet modes according to \eqref{eq:dispersion}. By virtue of the normalisation adopted, the free-stream acoustic waves are simply defined by $k_a^\pm=\pm2\pi StM_a$.

\subsubsection{Finite-thickness flow model}
\label{subsubsec:rayleigh}
Writing the LEE equations in terms of the pressure, the compressible Rayleigh equation

{\small
\begin{equation}
\frac{\partial^2\hat{p}}{\partial r^2}+\left(\frac{1}{r}-\frac{2k}{\overline{u}_xk-\omega}\frac{\partial\overline{u}_x}{\partial r}-\frac{\gamma -1}{\gamma\overline{\rho}}\frac{\partial\overline{\rho}}{\partial r}+\frac{1}{\gamma\overline{T}}\frac{\partial\overline{T}}{\partial r}\right)\frac{\partial\hat{p}}{\partial r}-\left(k^2+\frac{m^2}{r^2}-\frac{\left(\overline{u}_xk-\omega\right)^2}{\left(\gamma -1\right)\overline{T}}\right)\hat{p}=0
\label{eq:Rayleigh}
\end{equation}}

\noindent is obtained, where $\gamma$ is the specific heat ratio for a perfect gas. The solution of the linear stability problem is obtained specifying a real or complex frequency $\omega$ and solving the resulting augmented eigenvalue problem $k=k\left(\omega\right)$, with $\hat{p}\left(r\right)$ the associated pressure eigenfunction. The eigenvalue problem is solved numerically by discretizing \eqref{eq:Rayleigh} in the radial direction using Chebyshev polynomials. A mapping function is used to non-uniformly distribute the grid points such that they are dense in the region of shear \citep{trefethen2000spectral}.

With the aim of keeping the model as simple as possible and exploring the impact of the shear-layer thickness, we use the hyperbolic-tangent velocity profile \citep{lesshafft2007linear}

\begin{equation}
\overline{u}_x\left(r\right)=\frac{1}{2}M_a\left(1+\tanh\left(\frac{R}{4\theta_R}\left(\frac{R}{r}-\frac{r}{R}\right)\right)\right)\mathrm{,}
\label{eq:hyperbolic}
\end{equation}

\noindent where $\theta_R$ is the shear-layer momentum thickness and $R$ is the nozzle radius.

Different values of $R/\theta_R$ are considered to assess the impact of the shear-layer thickness on screech generation and, specifically, on the frequency range of existence of propagative guided modes. Consistent with results obtained from experimental particle image velocimetry (PIV) measurements (see appendix \ref{sec:PIV}), we select a shear layer with $R/\theta_R=10$.

\subsection{Resonance conditions}
\label{subsec:res_cond}
Resonance conditions are obtained by assuming that upstream- and downstream-travelling waves exchange energy at the nozzle exit and at the $s$-th shock-cell location, where an interaction between the K-H instability wave and a representative source point in the shock-cell pattern occurs. Following \cite{landau1958statistical}, the condition to be satisfied for resonance to occur is

\begin{equation}
R_1R_2e^{i\Delta kL_s}=1\mathrm{,}
\label{eq:res_condition}
\end{equation}

\noindent where $R_1\left(St,M_j\right)\in \mathcal{C}$ and $R_2\left(St,M_j,L_s\right)\in \mathcal{C}$ are the reflection coefficients at the nozzle exit and shock-cell location, respectively, $\Delta k=k^+-k^-$ is the difference between the wavenumbers of the downstream- and upstream-travelling waves and $L_s$ is the distance between the nozzle exit and the $s$-th shock cell. Following \cite{jordan2018jet}, \eqref{eq:res_condition} can be re-written in terms of magnitude and phase constraints associated with the imaginary and real parts of the eigenvalues $k$, respectively,

\begin{subequations}
\begin{align}
& e^{\Delta k_i L_s}=\vert R_1R_2\vert\mathrm{,}\label{eq:magnitude}\\
& \Delta k_r L_s +\phi =2p\pi\mathrm{,}\label{eq:phase}
\end{align}
\end{subequations}

\noindent where $\Delta k_i$ and $\Delta k_r$ are the imaginary and real parts of $\Delta k$, $\phi=\angle R_1R_2$ is the phase of the reflection-coefficient product and $p$ is an integer associated with the sum of oscillation peaks of the downstream- and upstream-travelling waves within the distance $L_s$. We show in appendix \ref{sec:wave} the equivalence between the phase criterion in \eqref{eq:phase} and the wave-interaction model first proposed by \cite{tam1982shock}.

Following our previous results \citep{mancinelli2019screech} and in agreement with observations from experiments and numerical simulations \citep{umeda2001sound, panda1999experimental, gao2010multi, mercier2017experimental} that suggest stronger interaction in the vicinity of the fourth shock-cell, i.e. $s=4$, we here construct our model on the basis that this is the dominant equivalent source dictating the screech frequency selection. The distance $L_s$ between the nozzle exit and the $s$-th shock cell is given by

\begin{equation}
L_s\left(M_j\right)=L_1\left(\left(1-\alpha\right)s+\alpha\right)\mathrm{,}
\label{eq:Ls}
\end{equation}

\noindent where $L_1$ is the length of the first shock cell and $\alpha$ is the rate of decrease of the shock-cell length with the downstream distance, whose value was estimated as $0.06$ by \cite{harper1974noise}. The length of the first shock cell can be obtained using the V-S model and by carrying out a zero-frequency analysis. When $\omega\to 0$ the dispersion relation in \eqref{eq:dispersion} reduces to

\begin{equation}
I_m\left(\frac{\gamma_i}{2}\right)=0\mathrm{,}
\label{eq:disp_0}
\end{equation}

\noindent which, exploiting the properties of the Bessel functions, can be written as

\begin{equation}
J_m\left(\frac{i\gamma_i}{2}\right)=0\mathrm{.}
\label{eq:J_I}
\end{equation}

Since we intend to describe the mean-flow shock-cell structure of an axisymmetric jet, we consider $m=0$. The first zero of the Bessel function $J_0$ in \eqref{eq:J_I} occurs when

\begin{equation}
i\frac{\gamma_i}{2}=2.4048\mathrm{.}
\label{eq:J}
\end{equation}

Using \eqref{eq:gamma_i} to replace $\gamma_i$ in \eqref{eq:J}, we obtain the Prandtl-Pack model \citep{pack1950note} for the length of the first shock cell,

\begin{equation}
L_1\left(M_j\right)=\frac{\pi}{2.4048}\sqrt{M_j^2-1}\mathrm{.}
\label{eq:L1}
\end{equation}

The reflection coefficients $R_1$ and $R_2$ are unknown functions of frequency and flow conditions. Two modelling approaches may be pursued in which the frequency, $\omega$, is either real or complex, and in each approach the reflection-coefficient product is handled differently. Following \cite{mancinelli2019screech}, we first use the real-frequency analysis to show that the feedback process in the generation of A1 and A2 screech modes is underpinned by guided jet modes rather than free-stream acoustic waves considered in the closure mechanism proposed by \cite{powell1953mechanism} and most other studies since. We then explore both temperature and shear-layer thickness effects on screech. This allows us to identify the limitations of real-frequency-based models and explore improvements that may be obtained using the complex-frequency model.

\subsection{Neutral-mode model}
\label{subsec:real_analysis}
\begin{figure}
\centering
\subfigure[]{\includegraphics[scale=0.4]{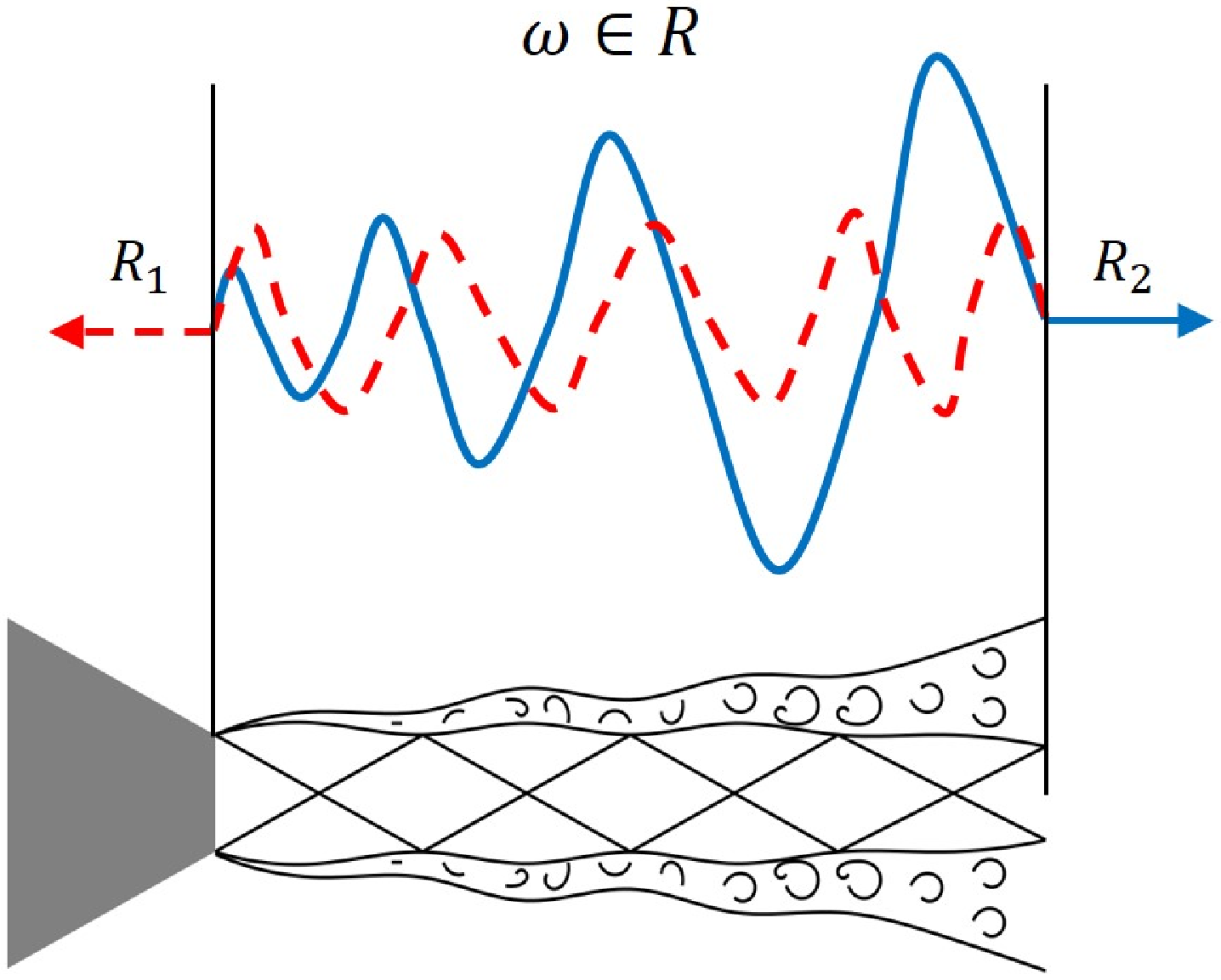}\label{fig:real_analysis}}$\;$
\subfigure[]{\includegraphics[scale=0.4]{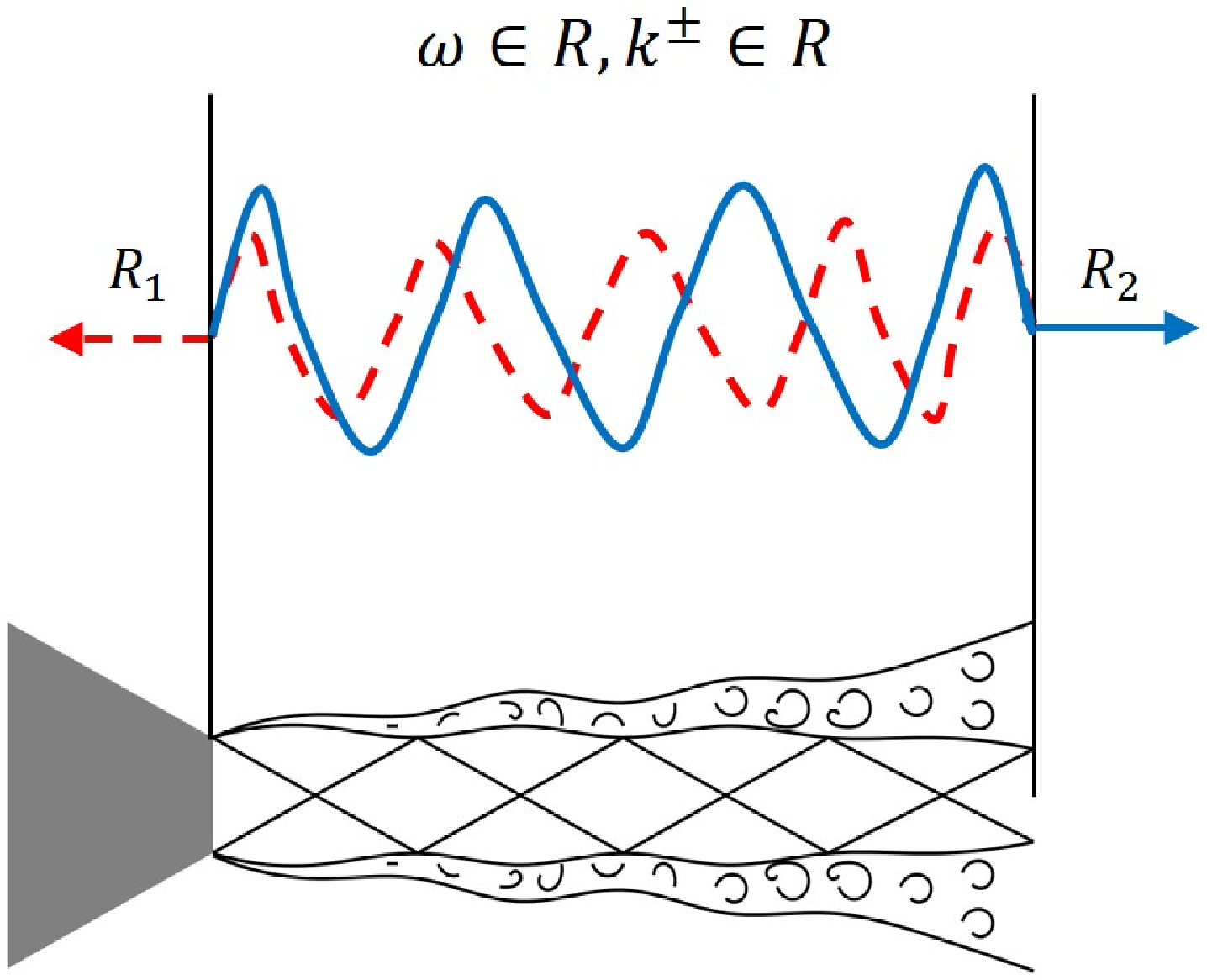}\label{fig:neutral}}
\caption{Schematic representation of the resonance loop: (a) real-frequency analysis, (b) neutral-mode model. Solid blue and dashed red lines refer to downstream- and upstream-travelling waves, respectively.}
\end{figure}

If the frequency is considered real, resonance involves a spatially-unstable downstream-travelling K-H mode and a neutrally-stable upstream-travelling wave: $\omega\in\mathcal{R}$, $k^+\in\mathcal{C}$, and $k^-\in\mathcal{R}$. The reflection coefficients $R_1$ and $R_2$ at the nozzle exit and shock-cell location, respectively, are such that resonance involves tones that neither grow nor decay in time. A schematic depiction of this scenario is provided in figure \ref{fig:real_analysis}.

A further widely used simplification involves assuming that both the downstream- and upstream-travelling waves are neutrally stable: $k^+$ and $k^-\in\mathcal{R}$ (see schematic representation in figure \ref{fig:neutral}). Since $\Delta k_i$ is then identically zero, this allows us to neglect the magnitude constraint of \eqref{eq:magnitude}. Resonance-frequency prediction can be obtained either from the dispersion relation \eqref{eq:dispersion} or from the Rayleigh equation \eqref{eq:Rayleigh} by specifying real values of $\omega$, computing $k$ and then imposing the phase constraint \eqref{eq:phase} to obtain the resonance frequency. Given that the phase of the reflection-coefficient product is an unknown, we follow \cite{jordan2018jet} and \cite{mancinelli2019screech} by exploring different phase values: $\phi=0$, $-\pi/4$, $-\pi/2$ and $-\pi$, which lead to the following resonance criteria,

\begin{subequations}
\begin{align}
& \phi=0\;\Longrightarrow\; Re\left[k^+-k^-\right]=\Delta k_r=\frac{2p\pi}{L_s}\label{eq:in_phase}\mathrm{,}\\
& \phi=-\pi/4\;\Longrightarrow\; Re\left[k^+-k^-\right]=\Delta k_r=\frac{\left(2p+1/4\right)\pi}{L_s}\label{eq:45_phase}\mathrm{,}\\
& \phi=-\pi/2\;\Longrightarrow\; Re\left[k^+-k^-\right]=\Delta k_r=\frac{\left(2p+1/2\right)\pi}{L_s}\label{eq:quad_phase}\mathrm{,}\\
& \phi=-\pi\;\Longrightarrow\; Re\left[k^+-k^-\right]=\Delta k_r=\frac{\left(2p+1\right)\pi}{L_s}\label{eq:out_phase}\mathrm{.}
\end{align}
\label{eq:phase_tot}
\end{subequations}

Predictions obtained using $\phi=-\pi/4$ \eqref{eq:45_phase} and in-phase reflection conditions \eqref{eq:in_phase} provide best agreement with the experimental data when using the vortex-sheet and finite-thickness dispersion relations, respectively. This discrepancy in phase between the two flow models is expected since the phase speed of the K-H mode changes significantly between the vortex sheet and the finite-thickness model \citep{michalke1984survey}.

\begin{figure}
\centering
\subfigure[]{\includegraphics[scale=0.26]{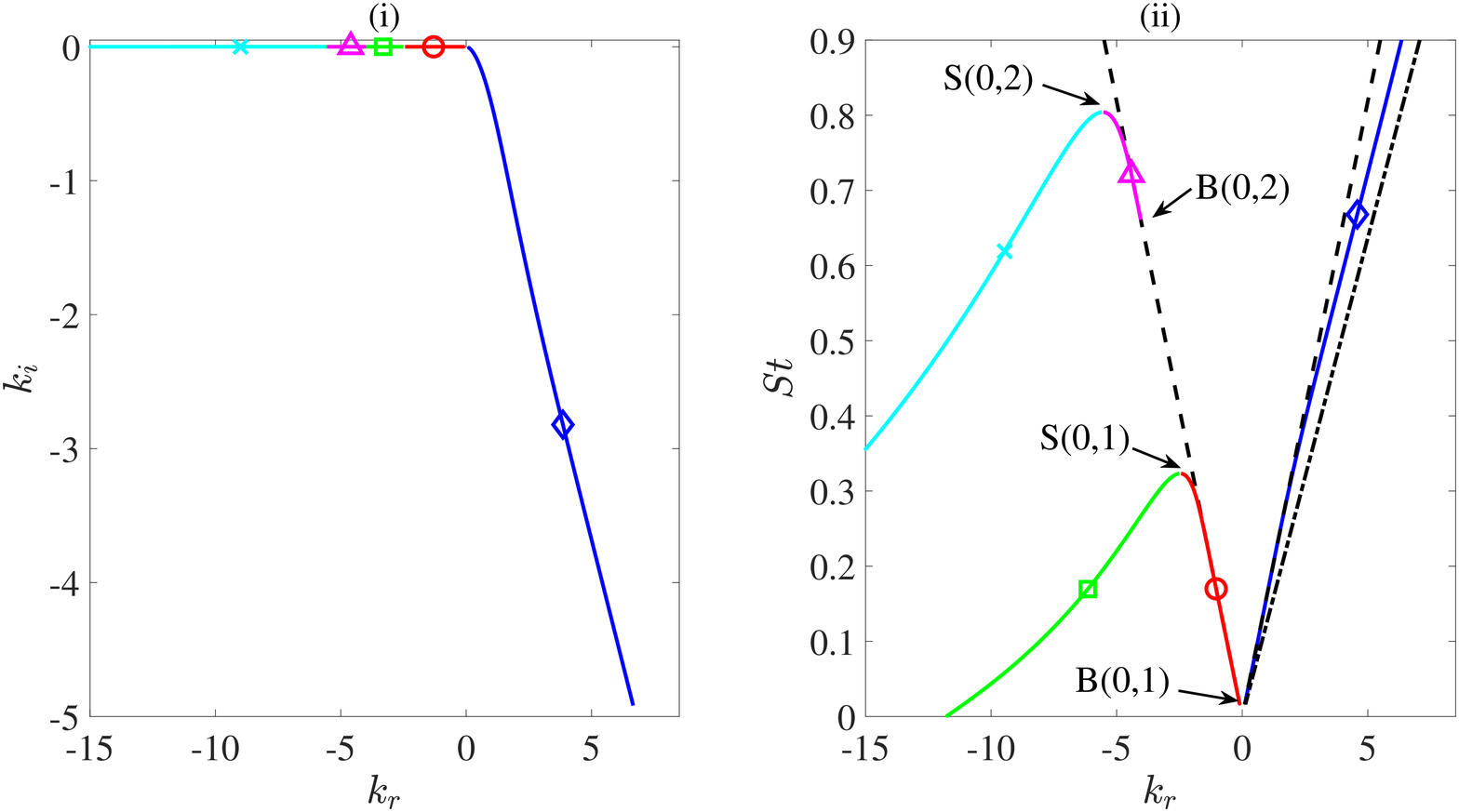}\label{fig:eig_VS}}
\subfigure[]{\includegraphics[scale=0.26]{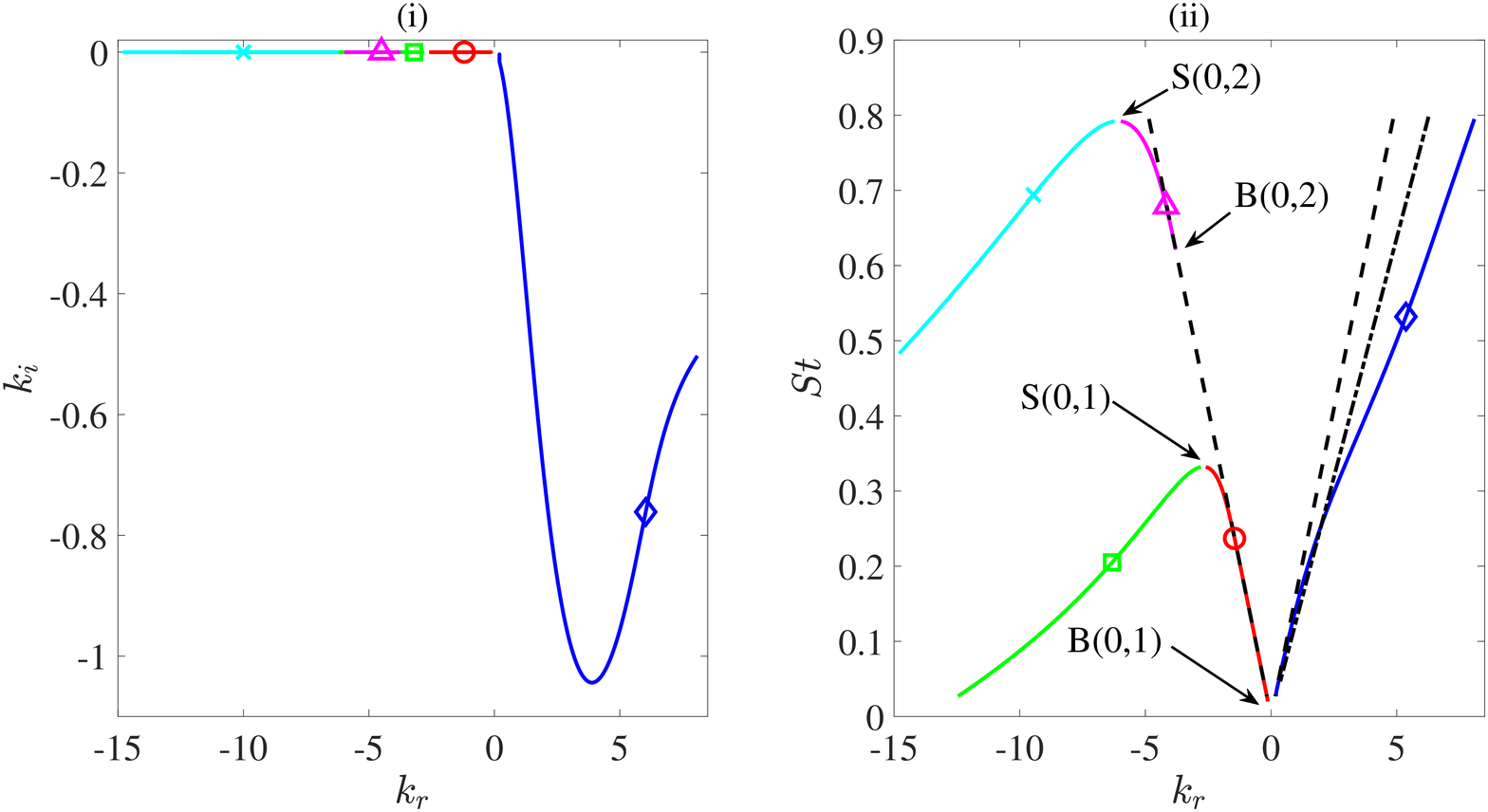}\label{fig:eig_RE}}
\caption{Eigenvalues $k_{KH}^+$, $k_p^\pm$ and $k_a^\pm$ in the case of a real-frequency analysis for azimuthal mode $m=0$, cold jet and fully expanded jet Mach number $M_j=1.08$: (a) vortex-sheet model, (b) finite-thickness model with $R/\theta_R=10$; (i) $k_r$-$k_i$ plane, (ii) $k_r$-$St$ plane. blue $\diamond$ corresponds to $k_{KH}^+$, red $\circ$ to $k_p^-$ for $n=1$, green $\Box$ to $k_p^+$ for $n=1$, magenta $\bigtriangleup$ to $k_p^-$ for $n=2$, cyan $\times$ to $k_p^+$ for $n=2$, dash-dotted black line to a non-dispersive approximation of K-H mode whose phase speed is $0.8\,U_j$, dashed black lines to $k_a^\pm$. The branch and saddle points of the guided jet modes for each pair of $\left(m,n\right)$ orders are indicated with letters $B$ and $S$, respectively.}
\label{fig:eigenvalues_real}
\end{figure}

Figure \ref{fig:eig_VS} shows the eigenvalue trajectories for azimuthal mode $m=0$, as a function of frequency, in the $k_r$-$k_i$ and $k_r$-$St$ planes for the V-S dispersion relation. Shown are the trajectories of the K-H and guided jet modes, and of the downstream- and upstream-travelling free-stream acoustic waves. The flow considered is cold and fully expanded, with temperature ratio $T=T_j/T_\infty\approx 0.81$ and jet Mach number $M_j=1.08$. A non-dispersive wave with a phase speed equal to $0.8\,U_j$ frequently adopted to approximate the K-H mode \citep{schmidt2017wavepackets} has been added to the plot. The $k_{KH}$ mode is slightly dispersive with phase speed greater than the usually adopted value of $0.8\,U_j$ as shown by the comparison with the constant phase speed line, $St=\left(0.8Uj\right)k$. We note the following features for the guided jet modes. They are characterised by a negative phase speed and, following \cite{tam1990theoretical} and \cite{towne2017acoustic}, we distinguish between guided jet modes with positive and negative group velocities. Only the $k_p^-$ modes may close the resonance loop. Unlike free-stream acoustic waves, they are dispersive, with subsonic phase speed very close to the value of the ambient speed of sound, particularly in the low-frequency part of the branch. They furthermore exist as propagative waves only in a well-defined Strouhal number range delimited by the branch and saddle points, $B\left(m,n\right)$ and $S\left(m,n\right)$, respectively \citep{tam1989three}. Outside this frequency band, the guided modes are evanescent and therefore not included in the modelling. For the first radial order, the branch point coincides with the origin in the $k_r$-$St$ plane \citep{tam1990theoretical}.

The trajectories of the K-H and guided jet modes using the finite-thickness dispersion relation with $R/\theta_R=10$ for the same flow conditions considered above are shown in figure \ref{fig:eig_RE}. The $k_{KH}$ wave shows a higher dispersive nature with a much lower phase speed than that computed using the vortex sheet, as discussed above. We also note that, contrary to the VS, the amplitude of the growth rate of the $k_{KH}$ mode is not monotonically increasing with the frequency.

\begin{figure}
\centering
\includegraphics[scale=0.25]{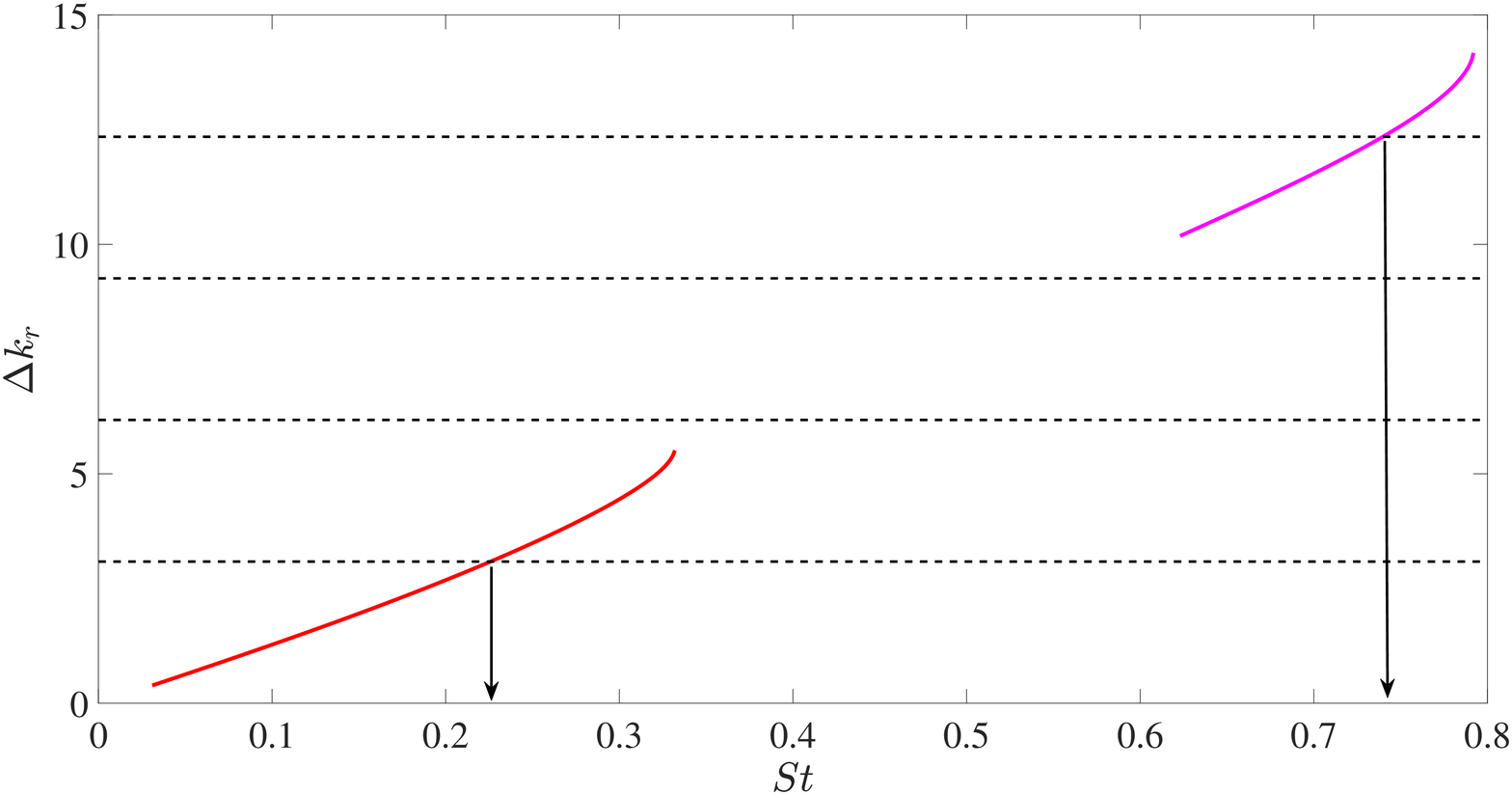}
\caption{$\Delta k_r$ between $k_{KH}^+$ and $k_p^-$ modes computed using the finite-thickness model and identification of the resonance frequency in the case of neutral-mode assumption for azimuthal mode $m=0$, $M_j=1.08$ and $T\approx 0.81$. Solid red line refers to $k_p^-$ for $n=1$, dash-dotted magenta line to $k_p^-$ for $n=2$, horizontal dashed black lines to resonance criteria in the case of in-phase reflection conditions \eqref{eq:in_phase}.}
\label{fig:res_cond_real}
\end{figure}

Figure \ref{fig:res_cond_real} shows the resonance frequency selection in the case of a resonance between $k_{KH}^+$ and $k_p^-$ modes of first and second radial orders for azimuthal mode $m=0$, $M_j=1.08$, $T_j/T_\infty\approx 0.81$ and in-phase reflection conditions. For the sake of brevity, we only report the results obtained using the finite-thickness dispersion relation. Resonance can only exist in a frequency range where both downstream- and upstream-travelling waves exist and are propagative. It is thus clear that when $k_p^-$ modes are considered, eligible resonance frequencies lie in the band delimited by the branch and saddle points. The resonance frequency is found by imposing the phase constraint, which implies computing the intersection between the $\Delta k_r$ and the value on the right hand-side of the equation of the in-phase resonance criterion \eqref{eq:in_phase}.

\subsection{Complex-mode model}
\label{subsec:complex_analysis}
\begin{figure}
\centering
\includegraphics[scale=0.35]{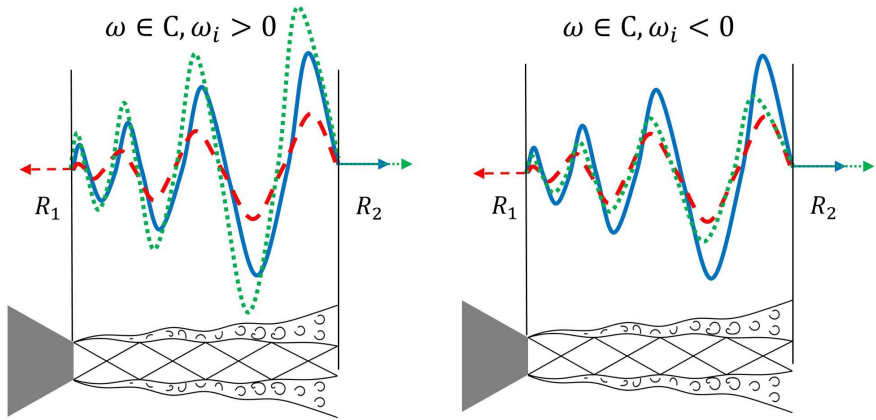}
\caption{Schematic representation of the resonance loop in the complex-frequency analysis. Solid blue lines refer to the downstream-travelling wave, dashed red lines to the upstream-travelling wave and dash-dotted green lines to the fed downstream-travelling wave in a new cycle of the resonance loop.}
\label{fig:complex_analysis}
\end{figure}

In the complex-frequency analysis both $k^+$ and $k^-\in\mathcal{C}$. The screech loop involves an unstable downstream-travelling K-H mode and an evanescent upstream-travelling wave. The complex-frequency analysis provides, as an additional result, information on the temporal amplification of the resonance loop: for $\omega_i>0$ the screech loop is amplified in time and resonance is sustained; for $\omega_i<0$ the screech loop is attenuated in time and resonance is damped. A schematic depiction of these scenarios is provided in figure \ref{fig:complex_analysis}.

Resonance-frequency prediction involves finding triplets $\left[k^+,k^-,\omega\right]\in\mathcal{C}$ simultaneously satisfying either the dispersion relation \eqref{eq:dispersion} or the Rayleigh equation \eqref{eq:Rayleigh} and both the magnitude and phase constraints \eqref{eq:magnitude} and \eqref{eq:phase}, respectively. This implies knowledge of the reflection-coefficient product $R_1R_2$ as a function of frequency and jet Mach number. As this function is unknown, we first use the reflection-coefficient product as a parameter with a prescribed amplitude. We then identify a frequency-Mach-number-dependent model for the reflection-coefficient product. Prediction involves solving \eqref{eq:dispersion} or \eqref{eq:Rayleigh} and \eqref{eq:magnitude} to find complex frequencies and eigenvalues that satisfy both the vortex-sheet dispersion relation or the Rayleigh equation and the magnitude constraint. The phase constraint \eqref{eq:phase} is then used to select resonance frequencies. 

\begin{figure}
\centering
\includegraphics[scale=0.27]{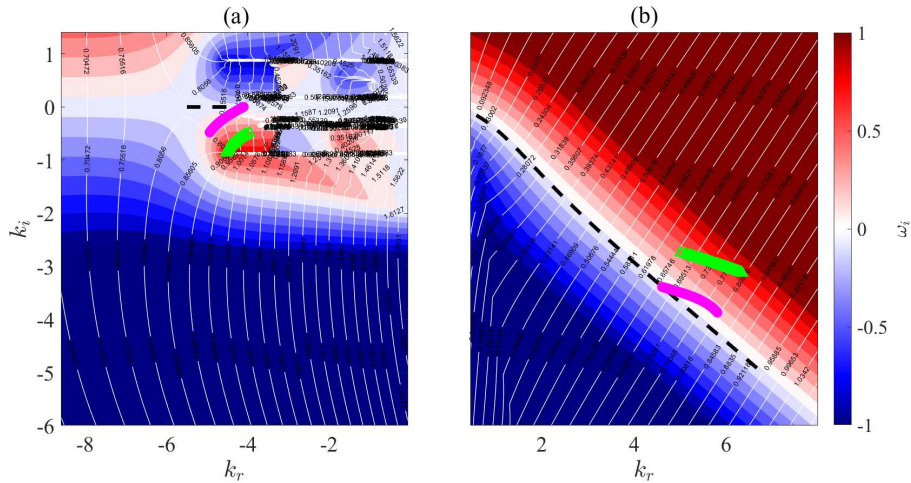}
\caption{Vortex-sheet solutions satisfying \eqref{eq:dispersion} in the case of complex-frequency analysis for azimuthal mode $m=0$, fully expanded jet Mach number $M_j=1.08$ and $T\approx 0.81$ for: (a) $k_p^-$ for $n=2$, (b) $k_{KH}^+$. The colour map shows $\omega_i$, whereas the white contour represents $St=\omega_r/2\pi M_a$. Dashed black lines represent the eigenvalues in the case of real-frequency analysis. Markers represent eigenvalues satisfying both the dispersion relation \eqref{eq:dispersion} and the magnitude constraint \eqref{eq:magnitude} for $s=4$ and different values of $\vert R_1R_2\vert$: magenta $\circ$ to $\vert R_1R_2\vert=10^{-3}$, green $\bigtriangleup$ to $\vert R_1R_2\vert=10^{-2}$.}
\label{fig:eigenvalues_complex}
\end{figure}

For the following discussion, we consider the eigenspectrum of the vortex sheet. This allows us to highlight certain features of the complex-mode model in comparison to the neutral-mode one. However, the complex-mode resonance conditions are used after to compute resonance frequency using the finite-thickness dispersion relation. Figure \ref{fig:eigenvalues_complex} shows the vortex-sheet solutions $\omega\left(k\right)$ for $k_{KH}^+$ and $k_p^-$ modes of second radial order for azimuthal mode $m=0$, Mach number $M_j=1.08$, $T\approx 0.81$ and with the fourth shock-cell location taken as downstream reflection point. Eigenvalues obtained from both real- and complex-frequency analyses are represented. Eigenvalues for $\omega\in\mathcal{C}$ are computed for two values of the reflection-coefficient product amplitude: $\vert R_1R_2\vert=10^{-3}$ and $10^{-2}$. As pointed out above, both downstream- and upstream-travelling waves are characterised by $k_i<0$ for $\omega\in\mathcal{C}$, implying that $k_{KH}^+$ and $k_p^-$ are, respectively, spatially unstable and spatially evanescent. Consistent with sustained resonance, the eigenvalues for $\omega\in\mathcal{C}$ move in the $k_r$-$k_i$ plane towards regions of positive imaginary frequency. Specifically, for a given value of the reflection-coefficient product, only $\omega_i\geq 0$ satisfies the amplitude condition in \eqref{eq:magnitude}. This feature of the complex-mode model allows us to identify frequency-Mach number combinations for which resonance is sustained.

\begin{figure}
\centering
\includegraphics[scale=0.25]{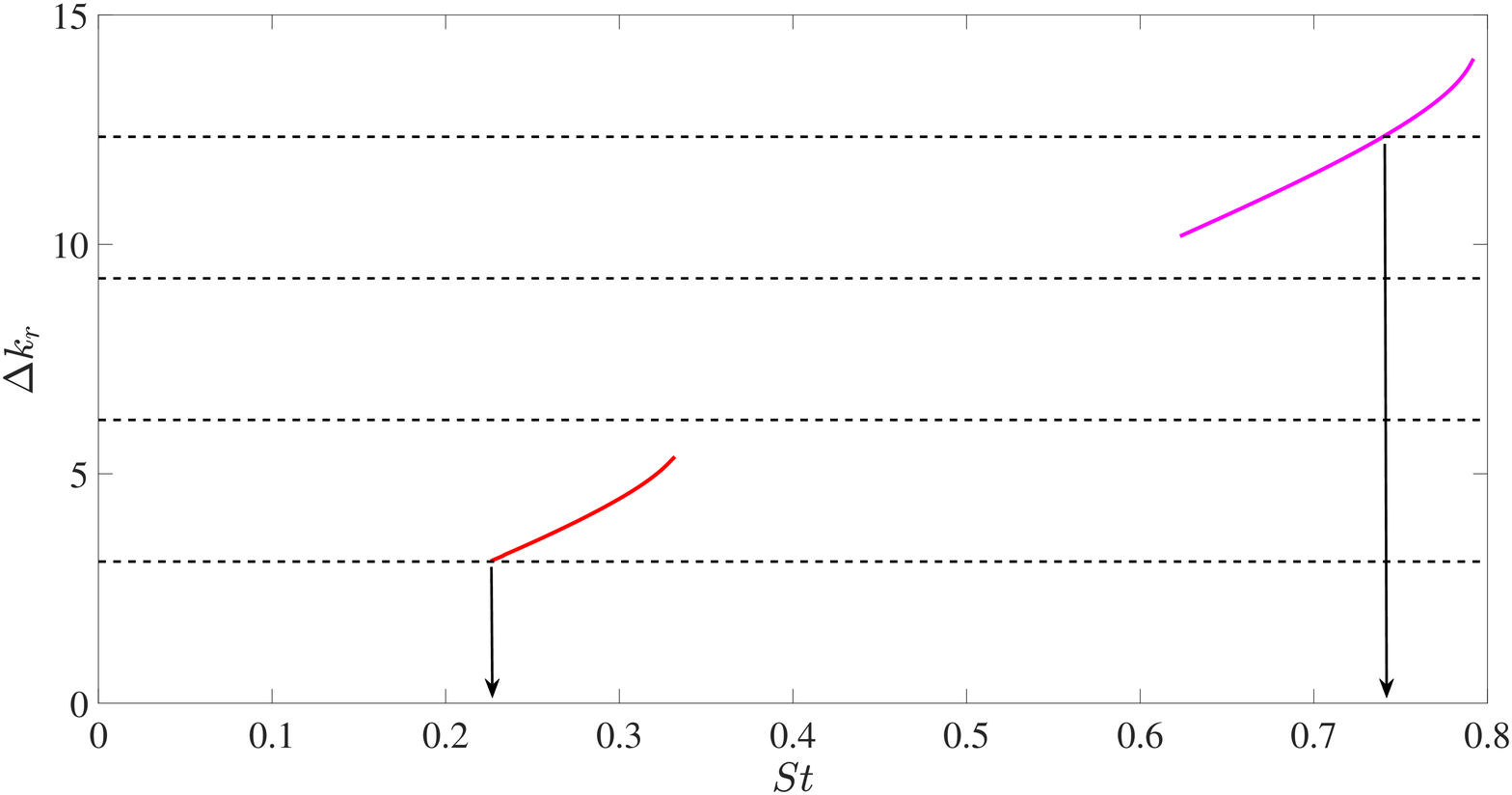}
\caption{$\Delta k_r$ between $k_{KH}^+$ and $k_p^-$ modes computed using the finite-thickness model and identification of the resonance frequency in the case of complex frequency for azimuthal mode $m=0$, $M_j=1.08$, $T\neq 1$ and magnitude of the reflection-coefficient product $\vert R_1R_2\vert=0.3$. Solid red line refers to $k_p^-$ for $n=1$, dash-dotted magenta line to $k_p^-$ for $n=2$, horizontal dashed black lines to in-phase resonance criteria \eqref{eq:in_phase}.}
\label{fig:res_cond_complex}
\end{figure}

Figure \ref{fig:res_cond_complex} illustrates the resonance-frequency selection when $\omega\in\mathcal{C}$ for azimuthal mode $m=0$, $M_j=1.08$ and $T\approx 0.81$. Consistent with what was found using finite-thickness neutral-mode model, best agreement is observed for in-phase reflection conditions \eqref{eq:in_phase}. In what follows, we compare model predictions with experiments.

\section{Experimental set-up}
\label{sec:setup}
The experimental test campaign was performed at the \textit{SUCR\'{E}} (SUpersoniC REsonance) jet-noise facility of the \textit{Prom\'{e}t\'{e}e} technological platform of the \textit{Institut Pprime} in Poitiers. The supersonic under-expanded jet issues from a simple convergent nozzle of diameter $D=10mm$ and nozzle-lip thickness of $0.3mm$. Experimental tests were carried out for a stagnation pressure range $p_0=\left[1.89,5.75\right]\cdot 10^5\,Pa$ with a corresponding fully expanded jet Mach number range $M_j=\left[1,1.8\right]$ and a nozzle diameter-based Reynolds number range $Re=U_jD/\nu=\left[2.86\cdot 10^5, 7.9\cdot 10^5\right]$. The jet facility includes a heating system to maintain a constant stagnation temperature $T_0=295K$. Fully-expanded conditions at the nozzle exit are computed from the stagnation conditions using the isentropic relations. The tests were performed with a very fine resolution $\Delta M_j=0.005$ for $M_j=\left[1,1.3\right]$ in order to capture the fine details of the Mach-number dependence of screech modes A1 and A2 and with a resolution $\Delta M_j=0.01$ for $M_j=\left[1.31,1.8\right]$.

\subsection{Acoustic measurements}
\label{subsec:p_measurements}
Pressure fluctuations were measured by GRAS 46BP microphones, whose frequency response is flat in the range $4\,Hz$-$70\,kHz$. Data were acquired by a National Instruments PXIe-1071 acquisition card with a sampling frequency of $200\,kHz$, which provides a maximum resolved Strouhal number range $\left[2, 3.2\right]$, well above the $St$ of interest in this paper. The acquisition time was set to $30\,s$, which is six orders of magnitude larger than the longest convective time, thus ensuring statistical convergence of the quantities presented in the paper. An azimuthal array of six microphones was placed in the nozzle-exit plane at a radial distance $r/D=1$ from the jet centreline. It was thus possible to resolve the most energetic azimuthal Fourier modes: $m=0,\pm 1,\pm 2$. The accuracy of the azimuthal Fourier decomposition at the tone frequencies was checked by computing the coherence between neighbouring microphones \citep{cavalieri2012axisymmetric}. Additional details on the jet facility and a schematic depiction of the experimental set-up and microphone disposition can be found in \cite{mancinelli2019screech} and \cite{mancinelli2019reflection}.

\subsection{Particle Image Velocimetry measurements}
\label{subsec:PIV_measurements}
PIV measurements were performed for several jet Mach numbers $M_j=1.08$, $1.12$, $1.16$, $1.22$, $1.3$, $1.43$, $1.47$ and $1.55$ in order to provide a description of the mean flow to inform the finite-thickness model (see \S\ref{subsubsec:rayleigh} and appendix \ref{sec:PIV}). The flow was seeded using ondina oil particles before entering the stagnation chamber ensuring seeding homogeneity. The particles were illuminated by a $2\times 200\,mJ$ Nd-YAG laser and the images were recorded with a $4\,Mpix$ CCD camera equipped with a Sigma DG Macro $105\,mm$ allowing the measurement of the jet up to $x/D = 10$ in the downstream direction. The PIV image pairs were acquired at a sampling rate of $7.2\,Hz$ with a $\Delta t$ of $1\mu s$. For each configuration, a total of 10000 image pairs were acquired in order to obtain well converged statistics. The images were processed using LaVision's Davis 8.0 software using a multi-pass iterative correlation algorithm \citep{willert1991digital, soria1996investigation} starting with an interrogation area of $64\times 64$ pixels and finishing with $16\times 16$ pixels with deforming windows. The overlap between neighbouring interrogation windows was 50\%, leading to a resolution of about 2.375 vectors per millimeter (i.e., 55 vectors per jet diameter) in the measured field. At each correlation pass, a peak validation criterion was used: vectors were rejected if the correlation peak was lower than 0.3. This value was selected as the minimum acceptable value ensuring validation in the potential regions of the flow while rejecting most of the evident erroneous vectors. Outliers were then further detected and replaced using universal outlier detection \citep{westerweel2005universal}. Main results of the PIV measurements are given in appendix \ref{sec:PIV}.

\section{Results}
\label{sec:results}
\begin{figure}
\centering
\includegraphics[scale=0.25]{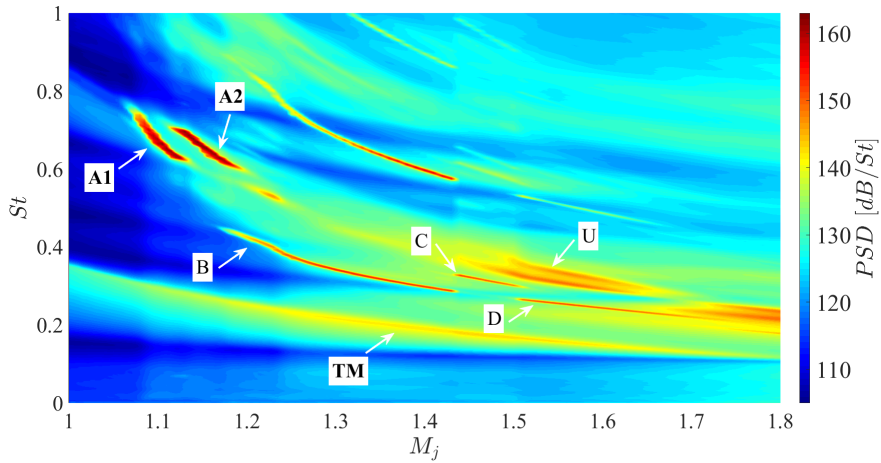}
\caption{$PSD$ map of mode $m=0$. Tones of interest, i.e., A1, A2 and TM, are labelled in bold.}
\label{fig:SPSL}
\end{figure}

Figure \ref{fig:SPSL} shows the Power Spectral Density ($PSD$) in $dB/St$ as a function of $St$ and $M_j$ for azimuthal mode $m=0$. The $PSD$ was computed using Welch's method \citep{welch1967use} with a Strouhal-number resolution $\left[1, 1.5\right]\cdot 10^{-3}$. The screech modes A1 and A2 are detected in the jet Mach number range $\left[1.07,1.23\right]$. A weaker low-frequency peak, which we hereinafter denote TM, is also detected for all Mach numbers. We furthermore detect the signature of asymmetric B, C and D modes in the mid-frequency range and of their harmonics for higher frequencies. We finally note that a weak signature of the U mode is detected just above the C mode. This screech stage was also observed by \cite{powell1992observations}, who speculated that it is the extension of the A2 mode. We show in appendix \ref{sec:bicoherence} that the appearance of B, C and D modes in the $PSD$ map of the axisymmetric component of the pressure is due to non-linear interaction between azimuthal modes $m=+1$ and $m=-1$. Our focus in what follows is the modelling of axisymmetric screech modes A1, A2 and TM.

\subsection{Neutral-mode analysis}
\label{subsec:results_real}

\subsubsection{Vortex-sheet model}
\label{subsubsec:results_VS}
\begin{figure}
\centering
\includegraphics[scale=0.27]{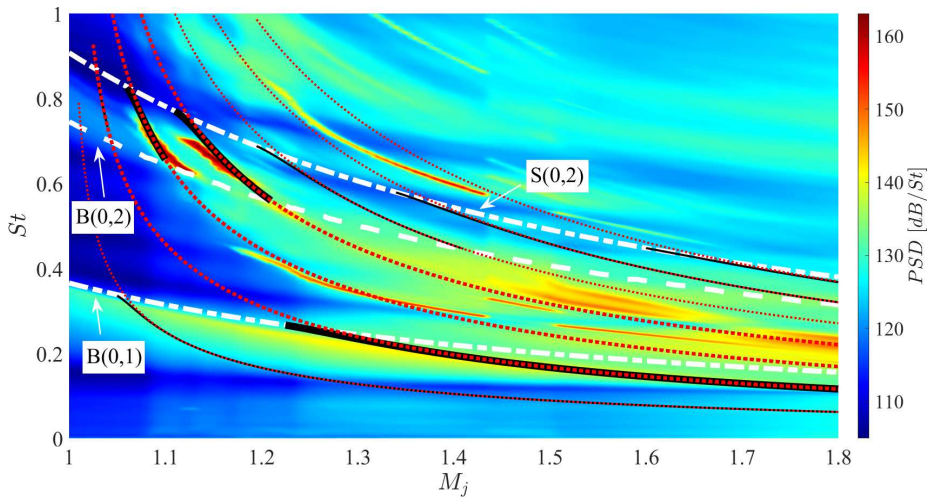}
\caption{$PSD$ map of mode $m=0$ for a cold jet and predictions from neutral-mode analysis using the vortex sheet. Dotted red lines refer to predictions obtained using $k_a^-$ waves, solid black lines to predictions obtained using $k_p^-$ modes for $n=1,2$. Dashed and dash-dotted white lines correspond to the branch- and saddle-point tracks, respectively. Predictions for TM, A1 and A2 tones are represented with bold lines.}
\label{fig:SPSL_predictions_real}
\end{figure}

Figure \ref{fig:SPSL_predictions_real} shows the $PSD$ map for $m=0$ and resonance-frequency predictions obtained using the neutral-mode analysis and considering both free-stream acoustic waves and guided jet modes. The branch- and saddle-point tracks are also shown. Predictions are computed using the vortex sheet to describe the jet-wave dynamics and considering a cold jet. We point out that the jet-to-ambient temperature ratio $T=T_j/T_\infty$ varies with $M_j$ in the range $\approx \left[0.61, 0.83\right]$. As outlined in \S\ref{subsec:real_analysis}, best predictions are obtained in this case by imposing $\phi=-\pi/4$ as reflection condition \eqref{eq:out_phase}, where we let the integer $p$ vary in the range $p=1,...,7$. Predictions for TM, A1 and A2 screech modes are obtained for values of $p$ equal to 2, 3 and 4, respectively. A summary of the parameters used to obtain predictions is reported in table \ref{tab:parameters}. For models in which guided jet modes close the resonance loop, A1 and A2 screech modes are associated with guided modes of second radial order ($n = 2$), and the TM mode is due to guided modes of first radial order ($n = 1$). Cut-on and cut-off frequencies of screech tones are not captured by the free-stream-acoustic-wave-based resonance model which predict resonance frequencies for regions where no tones are measured. In contrast, the guided-mode-based model roughly captures the cut-on and cut-off screech frequencies through the definition of a $M_j$-$St$ region of eligible resonance frequencies delimited by the branch- and saddle-point tracks. Although much more accurate than the model based on $k_a^-$ waves, the guided-mode-based model shows a rough agreement with experimental data. Specifically, the low-frequency part of A1 and, above all, A2 modes is cut out of the $St$-band of existence of propagative $k_p^-$ modes and hence cannot be predicted by the model. We underline that the better agreement with experiments observed in \cite{mancinelli2019screech} is, indeed, related to the fact that we considered an isothermal jet in that work. As we show later, the good description of the experimental data that we achieve by considering a more realistic cold jet with finite thickness of the shear layer was obtained using a vortex sheet in isothermal conditions in \cite{mancinelli2019screech} because temperature and shear-layer-thickness effects coincidentally cancel out in the model.

The impact of the jet temperature ratio on the branch- and saddle-point locations is addressed in appendix \ref{sec:T_effects}. We explore in the next section the improvements in the experimental data description that can be achieved by considering a flow model which takes into account the finite thickness of the shear layer of the jet.

\begin{table}
\centering
\begin{tabular}{ccccccc}
Resonance tone & $m$ & $n$ of $k_p^-$ & $T$ & $s$ & $\phi$ & $p$\\
\hline
A1 & 0 & 2 & $\neq 1$ & 4 & $-\pi/4$ & 3\\
A2 & 0 & 2 & $\neq 1$ & 4 & $-\pi/4$ & 4\\
TM & 0 & 1 & $\neq 1$ & 4 & $-\pi/4$ & 2\\
\end{tabular}
\caption{Summary of the parameters adopted to make resonance-frequency predictions using the vortex sheet.}
\label{tab:parameters}
\end{table}

\subsubsection{Finite-thickness model}
\label{subsubsec:results_RE}
\begin{figure}
\centering
\includegraphics[scale=0.25]{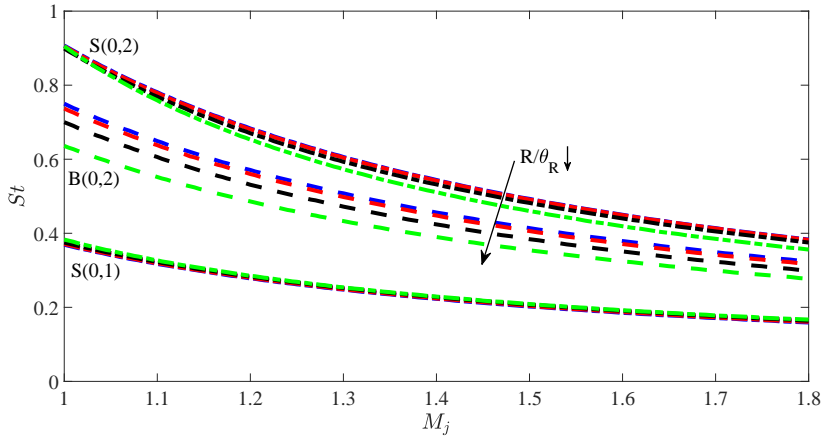}
\caption{Branch- and saddle-point tracks of guided modes of first and second radial order for a cold jet and for different shear-layer thickness values. Dash-dotted lines refer to the saddle point, dashed lines to the branch point. Blue lines correspond to shear layer with $R/\theta_R=30$, red lines to $R/\theta_R=20$, black lines to $R/\theta_R=10$, green lines to $R/\theta_R=5$.}
\label{fig:PSD_BS_Rtheta}
\end{figure}

We first explore the impact of the shear-layer thickness on the $St$-band of existence of propagative $k_p^-$ modes. No difference in the location of the branch and saddle points was observed between the vortex sheet and the finite-thickness dispersion relations for shear layers with $R/\theta_R\geq 30$, as large values of $R/\theta_R$ correspond to the V-S model. Figure \ref{fig:PSD_BS_Rtheta} shows the branch- and saddle-point tracks in the $M_j$-$St$ plane for different values of the shear-layer thickness. We choose thickness values with $R/\theta_R=30$, $20$, $10$ and $5$ and consider a cold jet in the model. The increase of the shear-layer thickness significantly changes the $St$-band of existence of the propagative guided mode of second radial order. Specifically, both the branch- and saddle-point tracks move to lower $St$ when the shear-layer thickness increases and this shift is not linearly progressive but becomes increasingly significant as $R/\theta_R$ moves away from the value of 30. The shift towards lower $St$ is more evident for $B\left(0,2\right)$ than for $S\left(0,2\right)$. Indeed, the shear-layer thickness effect on the saddle-point location appears to show a Mach-number dependence, with the shift towards lower $St$ much more significant for increasing $M_j$. The features just described result in a wider $St$-band of existence of $k_p^-$ mode for $n=2$ when the shear layer is thicker. No significant effect of the variation of $R/\theta_R$ is observed on the saddle point location of $k_p^-$ mode of the first radial order.

It is thus clear that the thickness of the shear layer is an important parameter in the definition of the $k_p^-$ modes and may strongly affect the accuracy of the screech-frequency prediction for modes A1 and A2. As detailed in appendix \ref{sec:PIV}, consistent with the results obtained from PIV data, we select $R/\theta_R=10$ from here on out.

\begin{figure}
\centering
\includegraphics[scale=0.25]{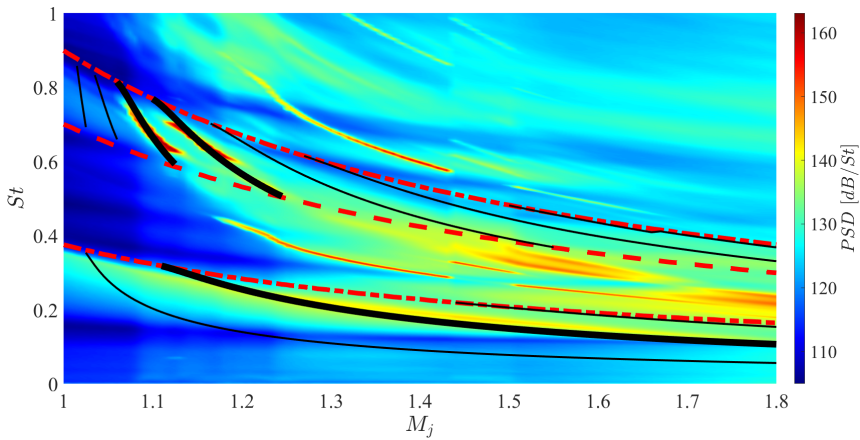}
\caption{$PSD$ map of mode $m=0$ and screech-frequency predictions obtained from neutral-mode analysis using $k_p^-$ modes for $n=1,2$ and considering a cold jet with $R/\theta_R=10$. Bold black lines correspond to predictions of modes A1, A2 and TM. Dashed and dash-dotted red lines correspond to the branch- and saddle-point tracks, respectively.}
\label{fig:SPSL_predictions_real_RE}
\end{figure}

Figure \ref{fig:SPSL_predictions_real_RE} shows the $PSD$ map of $m=0$ as a function of $St$ and $M_j$ and the screech-frequency predictions obtained using the guided modes as closure mechanism and considering a cold jet and a shear layer with $R/\theta_R=10$. Branch- and saddle-point tracks are also included in the plot. As outlined in \S\ref{subsec:real_analysis}, in-phase reflection conditions provided the best agreement with the experimental data. The list of parameters used to obtain screech-frequency predictions using a finite-thickness dispersion relation are summarised in table \ref{tab:parameters_RE}. In contrast to the V-S dispersion relation, both A1 and A2 screech modes are now entirely bounded by the branch- and saddle-point tracks. Predictions show a remarkable agreement with the experimental data for all screech modes A1, A2 and TM. Specifically, the accuracy of the prediction of the TM mode is strongly improved in comparison to that obtained using the vortex sheet: the mode is now accurately predicted in the range $M_j=\left[1.1,1.8\right]$, in contrast to the range $M_j=\left[1.25,1.8\right]$ obtained using the V-S model.

Nonetheless, as observed in the case of the vortex sheet, inaccuracies persist in the form of multiple tone predictions which do not correspond to measured tones: the model predicts tones for all the values of $p$ considered in \eqref{eq:in_phase}. This represents a limitation of the neutral-mode model. In the next section we explore the improvements that can be obtained using the complex-mode resonance conditions with the finite-thickness dispersion relation to describe the jet-wave dynamics.

\begin{table}
\centering
\begin{tabular}{cccccccc}
Resonance tone & $m$ & $n$ of $k_p^-$ & $R/\theta_R$ & $T$ & $s$ & $\phi$ & $p$\\
\hline
A1 & 0 & 2 & 10 & $\neq 1$ & 4 & $0$ & 4\\
A2 & 0 & 2 & 10 & $\neq 1$ & 4 & $0$ & 5\\
TM & 0 & 1 & 10 & $\neq 1$ & 4 & $0$ & 2\\
\end{tabular}
\caption{Summary of the parameters adopted to make resonance-frequency predictions using a finite-thickness model in the case of a cold jet.}
\label{tab:parameters_RE}
\end{table}

\subsection{Complex-mode analysis}
\label{subsec:results_complex}
\begin{figure}
\centering
\includegraphics[scale=0.29]{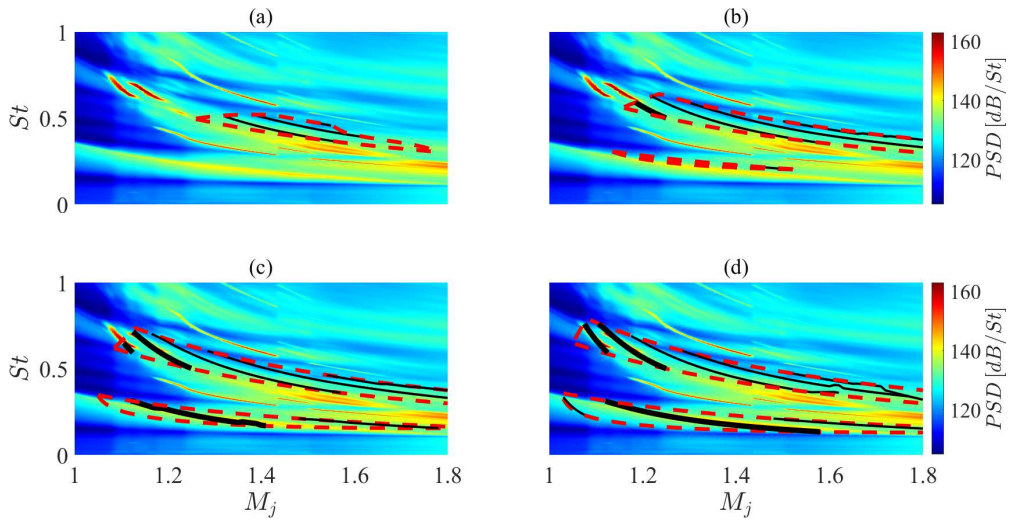}
\caption{$PSD$ maps of mode $m=0$ and predictions from complex-frequency analysis for different amplitude values of the reflection-coefficient product: (a) $\vert R_1R_2\vert=5\cdot 10^{-2}$, (b) $\vert R_1R_2\vert=10^{-1}$, (c) $\vert R_1R_2\vert=2\cdot 10^{-1}$, (d) $\vert R_1R_2\vert=3\cdot 10^{-1}$. The jet is considered cold with $R/\theta_R=10$. Predictions for TM, A1 and A2 screech tones are represented with bold black lines, spurious predictions with solid black lines. The region of allowable resonances for which $\omega_i\geq 0$ is marked by dashed red lines.}
\label{fig:SPSL_predictions_complex}
\end{figure}

Figure \ref{fig:SPSL_predictions_complex} shows $m=0$ $PSD$ maps and predictions obtained using the complex-mode resonance conditions. Predictions are obtained varying $p$ in eq. \eqref{eq:in_phase} from 1 to 7 and considering the same parameters listed in table \ref{tab:parameters_RE}. The threshold frequencies $\omega_r$ for which $\omega_i=0$ for all $M_j$ are also shown. These cut-off/cut-on frequencies define $M_j$-$St$ regions for which $\omega_i\geq 0$, where screech resonance may thus be sustained. Predictions can be obtained only within this region, and the extent of this region in the $M_j$-$St$ plane is a function of the amplitude of the reflection-coefficient product. We first explore four constant values of the reflection-coefficient product amplitude, $\vert R_1R_2\vert=5\cdot 10^{-2}$, $10^{-1}$, $2\cdot 10^{-1}$ and $3\cdot 10^{-1}$. We observe that for $\vert R_1R_2\vert=5\cdot 10^{-2}$ the $\omega_r\vert_{\omega_i=0}$ contours appear only for $k_p^-$ of second radial order and do not enclose any measured tones. As we increase the amplitude of the reflection-coefficient product, the region of positive imaginary frequencies extends to incorporate a broader $M_j$-$St$ region, allowable resonance regions appear and predictions of A2, A1 and TM modes gradually emerge. Specifically, the agreement between the resonance predictions of both A1 and A2 modes and the experimental data appears satisfactory for $\vert R_1R_2\vert=3\cdot 10^{-1}$, whereas the TM mode is well-predicted only in the range $M_j=\left[1.1,1.6\right]$. Spurious tone predictions are still observed for both $n=1$ and $n=2$ and these inaccuracies can be attributed to the crudeness of the reflection-coefficient product model.

The results shown above suggest a frequency-Mach-number-dependence of the amplitude of the reflection-coefficient product that must be considered in order to improve the accuracy of the resonance-prediction model.

\subsubsection{Real-frequency model for reflection-coefficient identification}
\label{subsubsec:RR_evaluation}
With the goal of improving the reflection-coefficient model, we use a variant of the complex-mode model to perform a data-driven identification of the reflection-coefficient product. This approach aims at finding a functional form for the amplitude of the reflection-coefficient product to guide the complex-mode resonance model. We consider K-H and guided jet waves for $\omega\in\mathcal{R}$, in which case $k_{KH}^+\in\mathcal{C}$ and $k_p^-\in\mathcal{R}$. For frequency-Mach number combinations for which tones are observed, we compute eigenvalues associated with these frequencies using the Rayleigh equation \eqref{eq:Rayleigh} and use the resonance condition in \eqref{eq:res_condition}, $R_1R_2e^{i\Delta kL_s}=1$, to calculate $R_1R_2\in\mathcal{C}$. This calculation provides the reflection-coefficient product for $\omega_i=0$, which corresponds to a feedback loop that neither grows nor decays in time. In the framework of the linear model, this amounts to the critical or minimum reflection-coefficient product necessary to sustain resonance and may be used as a first estimation to inform the full complex-mode model.

\begin{figure}
\centering
\includegraphics[scale=0.25]{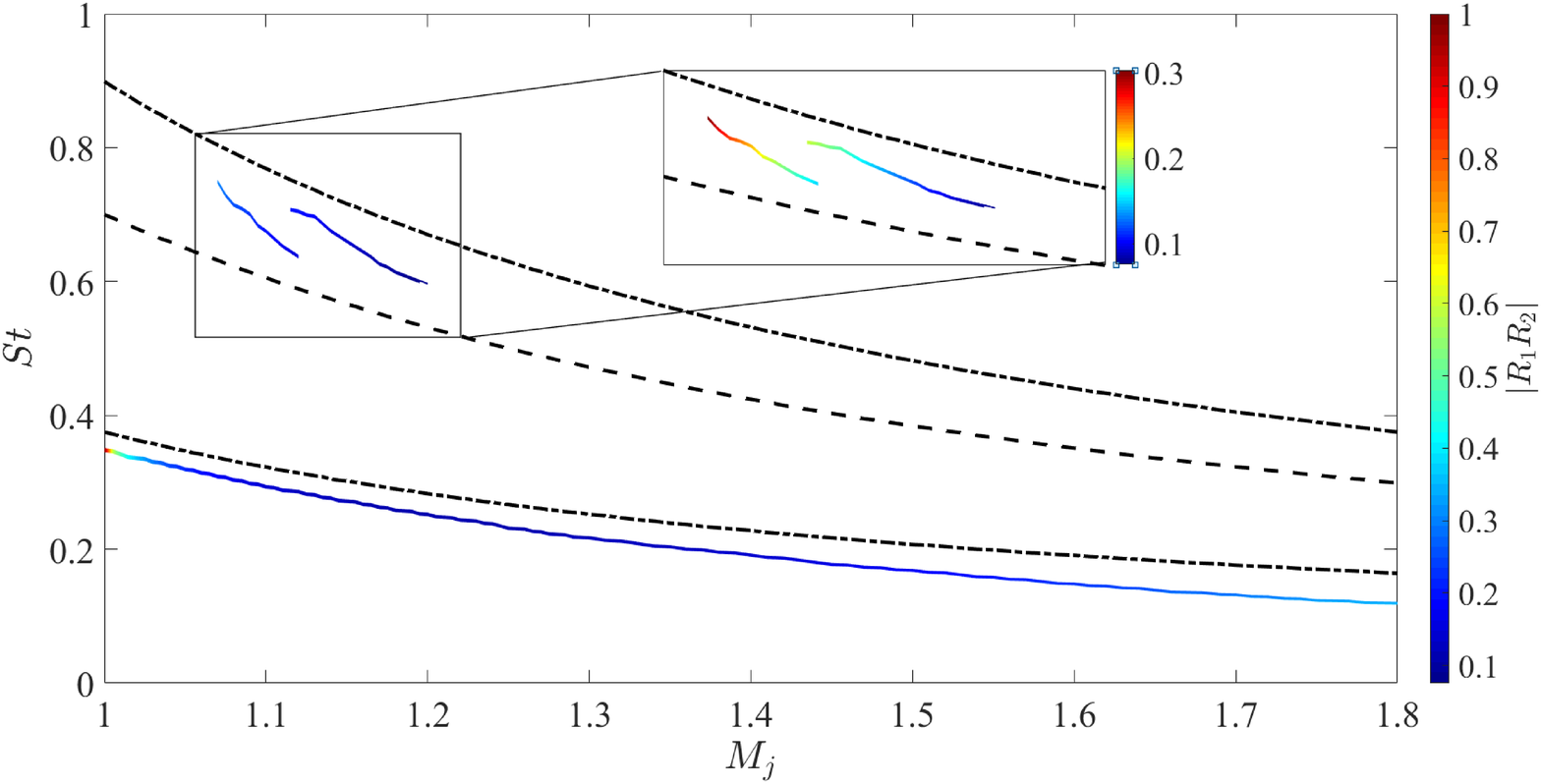}
\caption{Critical (or minimum) amplitude of the reflection-coefficient product computed from \eqref{eq:res_condition} considering $\omega\in\mathcal{R}$, $k_{KH}^+\in\mathcal{C}$ and $k_p^-\in\mathcal{R}$ for azimuthal mode $m=0$.}
\label{fig:RR_computed}
\end{figure}

\begin{figure}
\centering
\includegraphics[scale=0.25]{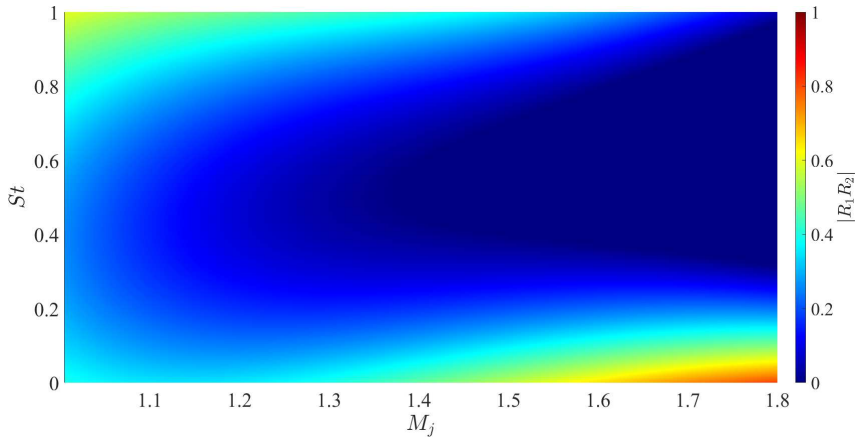}
\caption{Data-informed reflection-coefficient product amplitude as a function of $St$ and $M_j$, defined in \eqref{eq:RR_St_M} fit using the the critical $\vert R_1R_2\vert$ reported in figure \ref{fig:RR_computed}.}
\label{fig:RR_expStM}
\end{figure}

\begin{figure}
\centering
\includegraphics[scale=0.25]{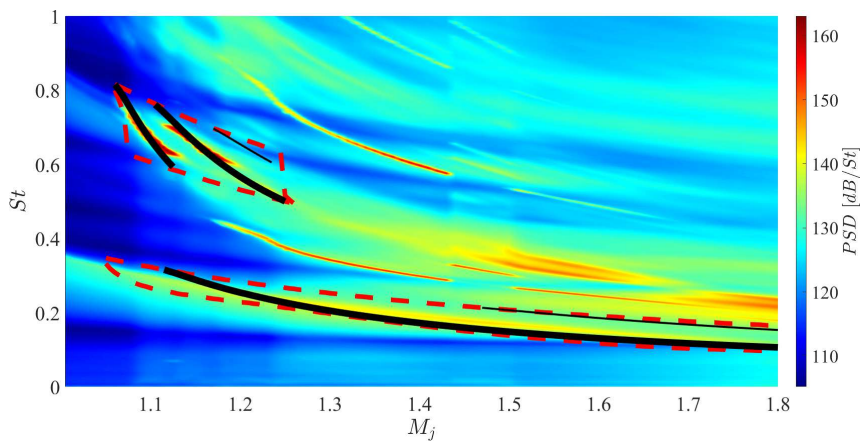}
\caption{$PSD$ map of mode $m=0$ and resonance-frequency predictions obtained using the data-informed $\vert R_1 R_2\vert$ function defined in \eqref{eq:RR_St_M} and shown in figure \ref{fig:RR_expStM}. Black lines correspond to predictions, dashed red lines delimit the region of allowable resonance frequencies, that is $M_j$-$St$ region where $\omega_i\geq 0$.}
\label{fig:SPSL_RR_expStM}
\end{figure}

Figure \ref{fig:RR_computed} shows the amplitude of the computed critical reflection-coefficient product for the measured tone frequencies. The result confirms the order of magnitude of $\vert R_1R_2\vert$ used earlier but does not suggest a straightforward functional form of the frequency-Mach-number dependence. With the aim of keeping the functional form as simple as possible, we explore polynomial-surface fits and select a polynomial surface of the third degree along $M_j$ and of second degree along $St$,

\begin{equation}
\vert R_1R_2\left(St,M_j\right)\vert=\sum_{i=0}^3\sum_{l=0}^2c_{il}M_j^iSt^l\qquad\text{with }i+l\leq 3\mathrm{,}
\label{eq:RR_St_M}
\end{equation} 

\noindent where the constants $c_{il}$ are parameters determined via a least-mean-square-based fit of the critical $\vert R_1R_2\vert$ values in figure \ref{fig:RR_computed}. The value of the coefficients $c_{il}$ is reported for the sake of completeness in table \ref{tab:fit}. The fitted model is shown in figure \ref{fig:RR_expStM} and the new predictions in figure \ref{fig:SPSL_RR_expStM}. Lines indicating $\omega_r\vert_{\omega_i=0}$ are also shown. We observe that the regions of allowable resonance frequencies for both $n=1,2$ are more tightly-fitted to the measured tones, the screech-frequency predictions are accurate and, more importantly, most of the spurious tone predictions have disappeared. Indeed, one spurious tone prediction is still observed for both $n=1$ and $n=2$. This could be likely ascribed to the limited amount of available data in the $M_j$-$St$ domain to build the functional form of the reflection-coefficient product amplitude (see figure \ref{fig:RR_computed}).

\begin{table}
\centering
\begin{tabular}{cccc}
\multicolumn{4}{c}{Value of coefficients $c_{il}$}\\
\hline 
i / l & 0 & 1 & 2\\
0 & 6.058 & 0.3351 & -1.275\\
1 & -12.42 & 0.2862 & 2.172\\
2 & 8.612 & -1.31 & \\
3 & -1.85 & &\\
\end{tabular}
\caption{Value of the polynomial fit coefficients in \eqref{eq:RR_St_M} used to build the functional form of the reflection-coefficient product.}
\label{tab:fit}
\end{table}

\section{Conclusions}
\label{sec:conclusions}
The resonance tones driven by axisymmetric azimuthal modes of a supersonic jet have been modelled. Attention was focused on the A1 and A2 screech modes and a low-frequency resonance peak, which we denoted TM, measured in the near pressure field of an under-expanded supersonic jet. The work follows our previous study \citep{mancinelli2019screech}, where we proposed a screech-frequency prediction model in which closure of the resonance loop is provided by upstream-travelling guided jet modes.

We here consider a resonance model that requires consideration of the upstream and downstream reflection coefficients, which appear in our model as the product $R_1R_2$. We treat this as a parameter.

We first consider a model in which the frequency is real and both the downstream- and upstream-travelling waves are neutrally stable. This simplification allows us to neglect the amplitude of the reflection coefficients at the nozzle exit and shock-cell location. Resonance-frequency predictions are achieved just by imposing the phase of the reflection-coefficient product. We explore temperature and shear-layer thickness effects on the band of existence of propagative guided modes and, hence, on screech generation. We detail the limitations of the vortex sheet in describing the jet dynamics to accurately make screech-frequency predictions. We then explore the improvements obtained considering a finite-thickness dispersion relation. Comparison with experimental data show how the predictions obtained considering a cold jet with a finite thickness value of the shear layer consistent with that obtained from Particle Image Velocimetry measurements are much more accurate. Inaccuracies persist, however, in the form of allowable resonance areas and spurious predictions for Strouhal and Mach number regions where no tones are observed.

We thus propose a more complete model in which both frequency and wavenumber are complex. In this framework, the downstream- and upstream-travelling waves are spatially unstable and evanescent, respectively. Both magnitude and phase reflection conditions must be considered in order to perform screech-frequency prediction. Positive values of the imaginary frequency identify regions of the $M_j$-$St$ plane where resonance may be sustained and predictions obtained. These regions depend on the amplitude of the reflection-coefficient product. We propose a frequency-Mach-number-dependent model for the reflection-coefficient product amplitude. The frequency-Mach-number-dependent functional form is obtained via a data-driven identification based on a real-frequency-based variant of the complex-mode model. The estimated functional form is then used to inform the full complex-mode model to predict screech tones. With this refinement, spurious tone predictions are almost completely removed and the model provides a more complete description of the experimental observations.

Future efforts should aim to develop an analytical model for the downstream and upstream reflection coefficients at the nozzle exit and shock-cell location without input from data; this work is currently underway by the authors. The complex-mode model could then provide screech-tone predictions.

\appendix

\section{Particle Image Velocimetry results}
\label{sec:PIV}
\begin{figure}
\centering
\includegraphics[scale=0.25]{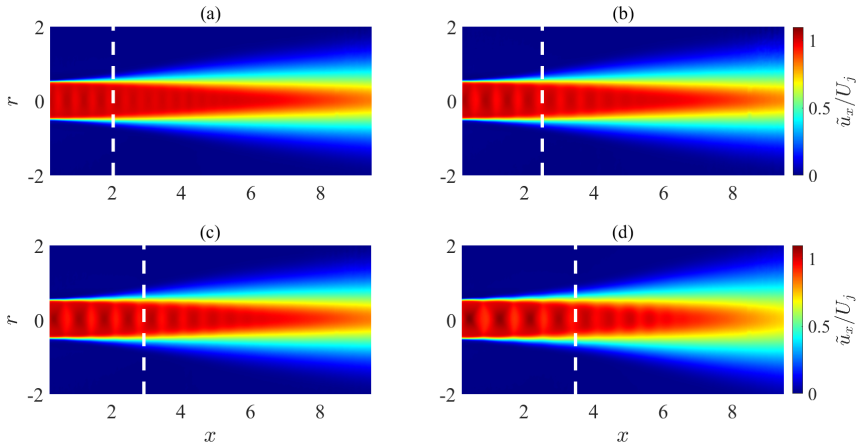}
\caption{Mean velocity field obtained from PIV measurements for different jet Mach numbers: (a) $M_j=1.08$, (b) $M_j=1.12$, (c) $M_j=1.16$, (d) $M_j=1.22$. White dashed lines indicate the position of the downstream reflection point, i.e., the fourth shock cell, according to \eqref{eq:Ls}.}
\label{fig:U_PIV}
\end{figure}

PIV measurements have been performed to provide a description of the mean velocity field. For the sake of brevity, we restrict the attention to the jet Mach numbers for which A1 and A2 screech tones are measured, that is $M_j=1.08$, $1.12$, $1.16$, $1.22$. Figure \ref{fig:U_PIV} shows the mean-velocity-field maps as a function of the axial and radial positions $x$ and $r$, respectively. The velocity values are normalised by the fully expanded jet velocity $U_j$, and the positions of the downstream reflection points, that is the fourth shock cell according to \eqref{eq:Ls}, are shown as well. We observe a good agreement between the shock reflection positions provided by \eqref{eq:Ls} and those measured by PIV. Specifically, the model slightly overestimates the axial location of the downstream reflection point but the discrepancy always stays under 10\%.

\begin{figure}
\centering
\includegraphics[scale=0.25]{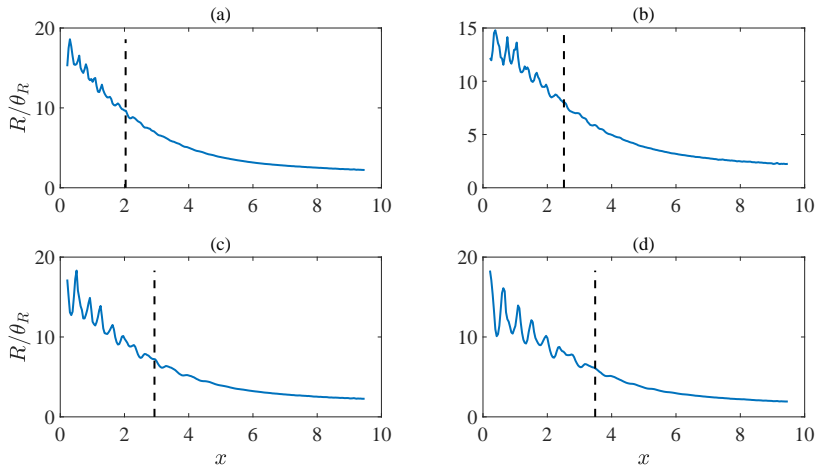}
\caption{Streamwise evolution of $R/\theta_R$, with $\theta_R$ the shear-layer momentum thickness, computed from the PIV mean velocity data for different jet Mach numbers: (a) $M_j=1.08$, (b) $M_j=1.12$, (c) $M_j=1.16$, (d) $M_j=1.22$. Black dashed lines indicate the position of the downstream reflection point, i.e., the fourth shock cell, according to \eqref{eq:Ls}.}
\label{fig:Rtheta_PIV}
\end{figure}

As outlined in \S\ref{subsubsec:rayleigh}, PIV data have been used to estimate the shear-layer momentum thickness and thus calculate the hyperbolic-tangent velocity profile \eqref{eq:hyperbolic} to inform the finite-thickness model for the jet-dynamics description \eqref{eq:Rayleigh}. Following \cite{michalke1984survey}, the shear-layer momentum thickness for a compressible jet is computed as follows,

\begin{equation}
\theta_R=\int\limits_0^\infty\frac{\tilde{\rho}\left(r\right)}{\rho^*}\frac{\tilde{u}_x\left(r\right)}{U^*}\left(1-\frac{\tilde{u}_x\left(r\right)}{U^*}\right)\,\mathrm{d}r\mathrm{,}
\label{eq:shear_thickness}
\end{equation}

\noindent where the $\sim$ is used to indicate the dimensional variables and the superscript $*$ was for the centreline values in the subsonic flow analysed by \cite{michalke1984survey}. We remind the reader that the centreline velocity corresponds to the maximum velocity value in the radial profile of a subsonic jet and this is not true for the shock-containing supersonic jet here analysed. We, hence, choose to use the superscript $*$ to denote the position for which the maximum mean velocity value is detected in the radial direction. The mean density profile $\tilde{\rho}\left(r\right)$ is obtained as the inverse of the temperature profile, $\tilde{\rho}\left(r\right)/\rho_j=\left(\tilde{T}\left(r\right)/T_j\right)^{-1}$, with the mean temperature computed from the mean velocity using the Crocco-Busemann relation \citep{michalke1984survey},

\begin{equation}
\frac{\tilde{T}\left(r\right)}{T_j}=\frac{T_\infty}{T_j} + \left(1-\frac{T_\infty}{T_j}\right)\frac{\tilde{u}_x\left(r\right)}{U_j} + \left(\gamma -1\right)M_j^2\frac{\tilde{u}_x\left(r\right)}{U_j}\frac{1}{2}\left(1-\frac{\tilde{u}_x\left(r\right)}{U_j}\right)\mathrm{,}
\label{eq:crocco}
\end{equation}

\noindent where $T_\infty$ is the ambient temperature and $U_j$ is the fully-expanded jet velocity. Figure \ref{fig:Rtheta_PIV} shows the streamwise evolution of the ratio $R/\theta_R$ for the jet Mach numbers $M_j=1.08$, $1.12$, $1.16$ and $1.22$. The location of the downstream reflection point is reported as well. As expected, the $R/\theta_R$ value decreases with increasing axial distance. Specifically, its value varies in the range $\left[6,18\right]$ from the nozzle exit to the downstream reflection location. With the aim of keeping the model as simple as possible, we choose to use a single analytical velocity profile \eqref{eq:hyperbolic} and we select $R/\theta_R=10$ as an approximate average value independent of both axial distance and jet Mach number.

\section{Relation with the wave-interaction model}
\label{sec:wave}
We here make explicit the relationship between the prediction model based on a resonance occurring between two fixed points at a distance $L_s$ (which is the perspective underlying our model),

\begin{equation}
\Delta k_rL_s+\phi=2\pi p\mathrm{,}
\label{eq:phase_appendix}
\end{equation}

\noindent and the wave-interaction model first proposed by \cite{tam1982shock},

\begin{equation}
k^+-k^-=k_s\mathrm{.}
\label{eq:wave_appendix}
\end{equation}

We make the hypothesis that the shock-cell structure can be represented as a distribution of equivalent point sources and consider both the simplified case of equi-spaced point sources and the more realistic case of irregularly-spaced point sources.

As mentioned in \S\ref{sec:intro}, for equi-spaced point sources, the shock-cell wavenumber can be simply expressed as $k_s=2\pi N_s/L_s$ and the wave-interaction model in \eqref{eq:wave_appendix} can be rearranged as

\begin{equation}
\Delta k_r L_s=2\pi N_s\mathrm{,}
\label{eq:Tam_wave_equi}
\end{equation}

\noindent where $\Delta k_r = k_r^+-k_r^-$. It is then straightforward to show that in order for \eqref{eq:Tam_wave_equi} and the resonance phase criterion in \eqref{eq:phase_appendix} to provide the same frequency prediction, the following relation must hold,

\begin{equation}
\frac{p-\frac{\phi}{2\pi}}{N_s}=1\mathrm{.}
\end{equation}

For irregular shock-cell spacings, substituting $\Delta k_r=k_s$ into \eqref{eq:phase_appendix} provides that the following relation must hold for the two models to provide the same resonance frequency,

\begin{equation}
k_s=\frac{2\pi p-\phi}{L_s}\mathrm{.}
\end{equation}

As shown by \cite{nogueira2021closure}, $k_s$ can be evaluated from the wavenumber spectrum of the shock-containing mean flow. Then, the wave-interaction model and the long-range resonance model we here present provide the same frequency predictions.

\section{Non-linear interaction between azimuthal modes}
\label{sec:bicoherence}
The appearance of the asymmetric B, C and D modes in the spectral content of mode $m=0$ due to non-linear interaction between modes $m=1$ and $m=-1$ is proved by computing the cross-bicoherence between these azimuthal modes. Non-linear quadratic interaction between azimuthal modes occurs when $m_1+m_{-1}-m_0=0$. Accordingly, the cross-bispectrum is computed as \citep{panickar2005nonlinear}

\begin{equation}
B_c\left(f, m_1,m_{-1},m_0\right)=\langle\hat{p}_1\left(f\right)\hat{p}_{-1}\left(f\right)\hat{p}_0^*\left(f\right)\rangle\mathrm{,}
\end{equation}

\noindent where $\hat{p}_m\left(f\right)$ is the Fourier transform of the $m^{th}$ azimuthal mode, the superscript $*$ indicates complex conjugate and the symbol $\langle\cdot\rangle$ denotes ensemble average between 2048-element segments of the whole time series, thus leading to the same Strouhal number resolution of the $PSD$ maps presented herein. A Hanning window and a 50\% overlap are used to perform the bispectrum computation. The cross-bicoherence is obtained by normalising the cross-bispectrum as

\begin{equation}
b_c^2\left(f, m_1,m_{-1},m_0\right)=\frac{\vert B_c\left(f, m_1,m_{-1},m_0\right)\vert^2}{\langle\vert\hat{p}_0\left(f\right)\vert^2\rangle\langle\vert\hat{p}_1\left(f\right)\hat{p}_{-1}\left(f\right)\vert^2\rangle}\mathrm{.}
\end{equation}

\begin{figure}
\centering
\includegraphics[scale=0.25]{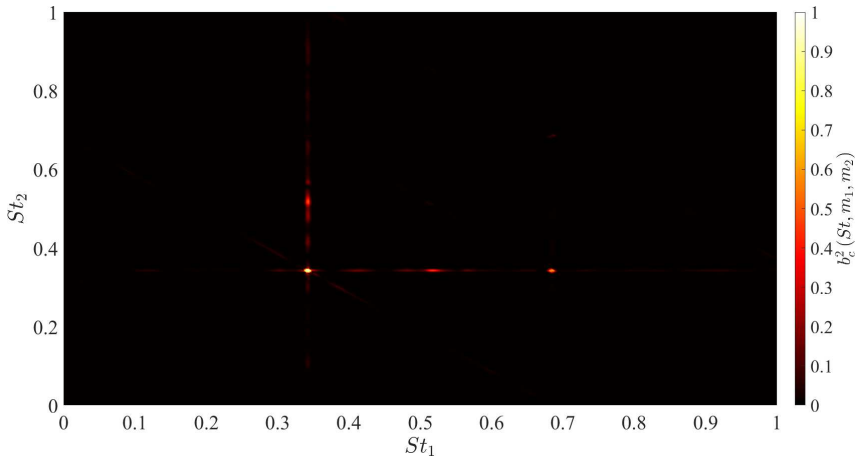}
\caption{Cross-bicoherence between azimuthal modes $m=1$, $m=-1$ and $m=0$ for jet Mach number $M_j=1.3$.}
\label{fig:bicoherence}
\end{figure}

Figure \ref{fig:bicoherence} shows the cross-bicoherence as a function of Strouhal number between azimuthal modes $m=1$, $m=-1$ and $m=0$ for $M_j=1.3$. For this jet Mach number, only screech mode B is active and its signature should be found in $m=1$ and $m=-1$ modes. We observe that a unitary bicoherence level is found for $St_1=St_2\approx 0.34$. This implies that a strong non-linear interaction occurs between modes $m=1$ and $m=-1$ at $St\approx 0.34$ leading to the appearance of an energy signature in mode $m=0$ for this frequency component, as confirmed by figure \ref{fig:SPSL}. We also observe low bicoherence levels along vertical and horizontal lines corresponding to the main screech frequency $St_1=St_2\approx 0.34$, where the subscript 1 and 2 refer to azimuthal modes $m=1$ and $m=-1$, respectively. These indicate negligible non-linear interactions between the main screech-frequency and all the other frequencies, with the exception of the interaction between the screech frequency and its first harmonic $St\approx 0.68$. The signature of this interaction is, indeed, detected by the appearance of the first harmonic of B screech mode in the $m=0$ $PSD$ map of figure \ref{fig:SPSL}.

\section{Temperature effects}
\label{sec:T_effects}
\begin{figure}
\centering
\includegraphics[scale=0.25]{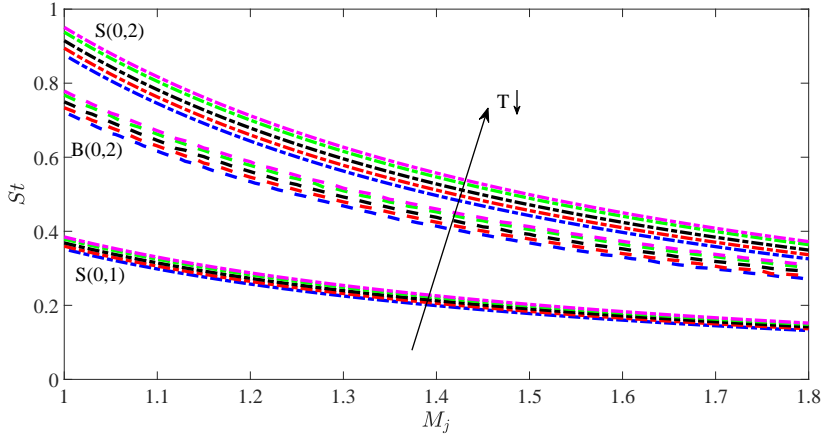}
\caption{Branch- and saddle-point tracks of guided jet modes in the $M_j$-$St$ plane for different jet-to-ambient temperature ratios $T$. Dash-dotted lines represent the saddle point and dashed lines represent the branch point. Blue lines refer to isothermal jet, i.e. $T=1$, red lines $T=0.9$, black lines $T=0.8$, green lines $T=0.7$, magenta lines to $T=0.65$.}
\label{fig:T_effect}
\end{figure}

We here explore the temperature effects on the branch- and saddle-point location of guided modes. For the sake of conciseness, we only report the results obtained using the vortex sheet for the jet-dynamics description and we underline that the same behaviour was observed in the finite-thickness model as well. Figure \ref{fig:T_effect} shows the branch- and saddle-point tracks of guided modes of $m=0$ and $n=1$, $2$ in the $M_j$-$St$ plane for jet-to-ambient temperature ratios $T=1$, $0.9$, $0.8$, $0.7$ and $0.65$. Both $B\left(0,n\right)$ and $S\left(0,n\right)$ tracks move to higher frequencies when the temperature decreases; this trend is more evident for the guided jet mode of second radial order than that of the first radial order. As outlined above, this behaviour explains why the vortex-sheet model in isothermal conditions that was used by \cite{mancinelli2019screech} provided a good description of the experimental data similar to that we here obtain using a finite-thickness model and considering a cold jet. Specifically, the shift towards lower frequencies of the branch- and saddle-point tracks due to the increase of temperature is coincidentally compensated by the shift towards higher frequencies due to the reduction of the shear-layer thickness.

\section*{Acknowledgments}
The authors acknowledge the financial support of EU and Nouvelle-Acquitaine region under the program CPER-FEDER. M.M. acknowledges the support of Centre National d'\'{E}tudes Spatiales (CNES) under a post-doctoral grant. Damien Eysseric is acknowledged for the support provided for the PIV measurements.

\section*{Declaration of Interests}
The authors report no conflict of interest.

\bibliographystyle{jfm}
\bibliography{biblio_rev}

\end{document}